\documentclass[11pt,a4paper]{article}
\usepackage{jcappub}
\usepackage{rotating} 
\usepackage{graphicx,epsfig}
\usepackage{amsmath}
\usepackage {amssymb}
\usepackage{subfigure}

\usepackage{relsize}

\newcommand{\be}{\begin{equation}}
\newcommand{\ee}{\end{equation}}
\newcommand{\bea}{\begin{eqnarray}}
\newcommand{\eea}{\end{eqnarray}}
\newcommand{\beaa}{\begin{eqnarray*}}
\newcommand{\eeaa}{\end{eqnarray*}}

\begin{document}

\title{Cosmology in time asymmetric extensions of general relativity}

\author[a]{Genly Leon}

\author[b,a]{Emmanuel N. Saridakis}

\affiliation[a]{Instituto de F\'{\i}sica, Pontificia
Universidad de Cat\'olica de Valpara\'{\i}so, Casilla 4950,
Valpara\'{\i}so, Chile}
\affiliation[b]{Physics Division, National Technical University of Athens,
15780 Zografou Campus,  Athens, Greece}

\emailAdd{genly.leon@ucv.cl}

\emailAdd{Emmanuel$_-$Saridakis@baylor.edu}

\abstract{ We investigate the cosmological behavior in a universe governed by time 
asymmetric extensions of general relativity, which is a novel modified gravity based on 
the addition of new, time-asymmetric, terms on the Hamiltonian framework, in a way that 
the algebra of constraints and local physics remain unchanged. Nevertheless, at 
cosmological scales these new terms  can have significant effects that can alter the 
universe evolution, both at early and late times, and the freedom in the choice of the 
involved modification function makes the scenario able to produce a huge class of 
cosmological behaviors. For basic ansatzes of modification, we perform a detailed 
dynamical analysis, extracting the stable late-time solutions. Amongst others, we find 
that the universe can result in dark-energy dominated, accelerating solutions, even in 
the absence of an explicit cosmological constant, in which the dark energy can be 
quintessence-like, phantom-like, or behave as an effective cosmological constant. 
Moreover, it can result to matter-domination, or to a Big Rip, or experience the 
sequence from matter to dark energy domination. Additionally, in the case of closed 
curvature, the universe may experience a cosmological bounce or turnaround, or even 
cyclic behavior. Finally, these scenarios can easily satisfy the observational and 
phenomenological requirements. Hence, time asymmetric cosmology  can be a good candidate 
for the description of the universe. }


\maketitle

\section{Introduction}
\label{Introduction}
 
The standard model of cosmology includes two accelerated phases of expansion, at early 
and late times respectively. Such a behavior cannot be obtained within the standard 
paradigm of physics, namely in the framework of general relativity and Standard Model of 
particles. Hence, additional degrees of freedom should be included in the picture. If 
these extra degrees of freedom are attributed to new, exotic ingredients of the universe 
content, then concerning late times one has the concept of dark energy (for reviews see  
\cite{Copeland:2006wr,Cai:2009zp}) and concerning early times the concept of inflaton 
field(s) (for reviews see \cite{Olive:1989nu,Bartolo:2004if}). On the other hand, if the 
extra degrees of freedom are of gravitational origin, then one obtains the paradigm of 
modified gravity (see \cite{Nojiri:2006ri,Capozziello:2011et} and references therein). 
The latter approach has the additional motivation of improving the UltraViolet 
behavior of gravity and alleviating the difficulties towards its quantization 
\cite{Stelle:1976gc,Biswas:2011ar}. Note that there are not strict boundaries between the 
above approaches, since one can partially or completely transform from one to the other, 
or construct theories where both extensions are imposed.

In the usual approach to gravitational modification one adds higher-order corrections to  
the Einstein-Hilbert action, like in $F(R)$ gravity 
\cite{DeFelice:2010aj,Nojiri:2010wj,Capozziello:2005ku,Amarzguioui:2005zq,Nojiri:2006gh}, 
in Gauss-Bonnet and $f(G)$ gravity \cite{Nojiri:2005jg,DeFelice:2008wz}, in Lovelock 
gravity \cite{Lovelock:1971yv,Deruelle:1989fj}, in Weyl gravity
\cite{Mannheim:1988dj,Flanagan:2006ra}, in 
Ho\v{r}ava-Lifshitz gravity
\cite{Horava:2008ih,Kiritsis:2009sh,Saridakis:2012ui}, in Galileon modifications 
\cite{Nicolis:2008in,Deffayet:2009wt,Deffayet:2009mn,Leon:2012mt},
in nonlinear massive gravity
\cite{deRham:2010kj,Hinterbichler:2011tt,deRham:2014zqa,Leon:2013qh} etc. A different 
class of gravitational modifications arise when one starts from the equivalent torsional 
formulation of gravity and add higher-order correction, like in $f(T)$ gravity  
\cite{Ben09,Linder:2010py,Chen:2010va,Cai:2011tc}, in $f(T,T_G)$ gravity
\cite{Kofinas:2014owa,Kofinas:2014aka,Kofinas:2014daa}, etc.
 
Recently, a new class of modified gravity was proposed \cite{Cortes:2015ola}. In 
particular, working in the Hamiltonian framework the authors constructed a theory 
that breaks the time reversal invariance of general relativity. Although the algebra
of constraints and local physics are unchanged, new terms appear at cosmological scales, 
that can alter the universe evolution, both at early and late times.

In the present work we are interesting in investigating in detail the cosmological 
implications of the above time asymmetric extensions of general relativity. In order to 
achieve this independently of the initial conditions and the specific universe evolution, 
we apply the dynamical systems method \cite{Coley:2003mj,Leon2011} which allows us to 
extract the global behavior of the scenario, bypassing the complexity of the involved 
equations. Indeed, due to the freedom in choosing the relevant extra modification 
function, the capabilities of the scenario are found to be huge. The plan of the work is 
the following: In section \ref{themodel} we present the time asymmetric extension of 
general relativity and we apply it in a cosmological framework. In section \ref{Dynanal} 
we perform a detailed dynamical analysis, extracting the stable late time solutions and 
the corresponding observables, and in section \ref{PhysImplic} we discuss their 
physical implications. Lastly, section \ref{Conclusions} is devoted to the conclusions.

\section{Time asymmetric extensions of general relativity and cosmology}
\label{themodel}

Let us briefly review the time asymmetric extension of general relativity
\cite{Cortes:2015ola}. In a first subsection we 
present the gravitational model itself, while in a second subsection we apply it in a 
cosmological framework.

\subsection{Time asymmetric extension of general relativity}

In this formulation one starts with the Hamiltonian form of 
general relativity  with a cosmological constant \cite{Arnowitt:1962hi}
\begin{equation}
S^{GR}= \int dt \int_\Sigma \left \{   \pi^{ab} \dot{g}_{ab} - N {\cal H}^{ADM} - 
N^a {\cal 
D}_a
\right \},
\end{equation}
where 
\begin{equation}
{\cal H}^{ADM} = -\frac{1}{G}  \sqrt{g} \left(R - 2 \Lambda\right) +\frac{G}{\sqrt{g}} 
\left(\pi^{ab} 
\pi_{ab} -\frac{1}{2} \pi^2\right) + {\cal H}^{\Psi} =0
\label{Hamiltonianconstraint}
\end{equation}
is the usual Hamiltonian constraint. In the above expressions $g_{ab}$ is the spatial 
metric, with $\pi^{ab}$ its canonical momenta and $\pi  = g_{ab}\pi^{ab}$ the 
corresponding trace, while $N$ and $N^a$ are the usual lapse and shift functions. In this 
formalism, the Hamiltonian constraint (\ref{Hamiltonianconstraint}), along with the 
diffeomorphism constraint
\begin{equation}
{\cal D}_a = D_b \pi_a^{b} + {\cal D}^{\Psi}_a =0,
\label{difconstraint}
\end{equation}
form a first class algebra, where the terms $ {\cal H}^{\Psi} $ and ${\cal 
D}^{\Psi}_a $ correspond to the matter content and $D_a$ is the covariant derivative.  
Obviously, the above expressions respect the time reversal symmetry 
\begin{subequations}
\begin{eqnarray}
\label{trev}
&& t  \rightarrow  -t   
\\
\label{grev}
&&g_{ab}   \rightarrow  g_{ab}
 \\
&&\pi^{ab}  \rightarrow  -\pi^{ab}.
\label{pirev}
\end{eqnarray}
\end{subequations}

In order to acquire well defined cosmological evolution equations one must use a gauge 
fixing, and it proves convenient to use the ``constant mean curvature gauge condition'' 
(CMC) \cite{Cortes:2015ola} 
\begin{equation}
\pi - \sqrt{g} <\pi > =0,
\label{CMCcond}
\end{equation}
where  $<\cdots >$ denotes the spatial average of a  density $\rho $ defined through
$<\rho > = \left(\int_\Sigma \rho\right)/\left(\int_\Sigma \sqrt{g}\right)$, with 
$V=\int_\Sigma \sqrt{g}$ the spatial volume. 
The CMC condition (\ref{CMCcond}) is a gauge fixing of the Hamiltonian constraint 
(\ref{Hamiltonianconstraint}), and thus they form a second class system. However, note 
that the CMC condition (\ref{CMCcond}) and the diffeomorphism constraint 
(\ref{difconstraint}) form a system of four first class constraints 
\cite{Gomes:2010fh,Gomes:2011zi,Gomes:2013naa}, as it is the case for the Hamiltonian 
constraint along with the diffeomorphism constraint. One can show that, restricting to 
constraints that are local in $g_{ab}$ and $\pi^{ab}$, there are no other pairs of 
systems of four first class constraints that one is the gauge fixing of the other, 
however 
one has the freedom to add a term linear in $\pi$ to the Hamiltonian constraint 
\cite{Gomes:2013naa}. This new term $\pi/L$, with $L$ the length-scale where this term 
becomes significant, breaks the time reversal symmetry (\ref{trev})-(\ref{pirev}), and 
this feature gave to the obtained gravitational modification the name  ``time asymmetric 
extension of general relativity''. One can extend the above extra, time-asymmetric, term 
of 
the Hamiltonian constraint, by assuming that the length-scale in which it becomes 
important is driven by a function of spatially averaged quantities, such as the spatial 
volume $V$. Hence, in summary, one can extend (\ref{Hamiltonianconstraint}) to a modified 
Hamiltonian constraint of the form \cite{Cortes:2015ola}
\begin{equation}
{\cal H}^{new} = - \frac{1}{G}  \sqrt{g} \left(R - 2 \Lambda\right) +\frac{G}{\sqrt{g}} 
\left(\pi^{ab} \pi_{ab} -\frac{1}{2} \pi^2\right)
+f(V) \pi  + {\cal H}^{\Psi}   =0,
\label{NewHamiltcontstr}
\end{equation}
where $f(V)$ is an arbitrary function of $V$. 
 
The above modification of the Hamiltonian constraint gives rise to a novel class of 
gravitational modifications. The new term leaves the constraint algebra and the local 
physical degrees of freedom unchanged \cite{Cortes:2015ola}. The only complexity comes 
from the fact that it affects the propagation of chiral fermions, since the left-handed 
spacetime connection $D_a \Psi_A$ does depend on $\pi^{ab}$. In order to handle this 
issue, one introduces the Ashtekar geometry \cite{Ashtekar:1986yd}, alongside the usual 
spacetime geometry characterized by the spacetime metric $g_{\mu \nu}$. Thus, although 
the 
gravitational effects and the propagation of photons are governed by the conventional 
spacetime geometry, the propagation of chiral fermions is determined by the  Ashtekar 
geometry which contains all the information of time irreversible behavior. Nevertheless, 
since in this work we are interested in the late-time background cosmological evolution, 
in which the matter sector is effectively described by a perfect fluid, and where 
radiation (a part of which is composed by chiral fermions) is negligible, in the 
following we do not discuss the above issue in more details. Hence, the time asymmetric 
modified gravity that we focus in this work is characterized by the action 
\begin{equation}
S= \int dt \int_\Sigma \left \{   \pi^{ab} \dot{g}_{ab} - N {\cal H}^{new} - 
N^a {\cal 
D}_a
\right \},
\label{newaction}
\end{equation}
where ${\cal H}^{new}$ is given by (\ref{NewHamiltcontstr}).

\subsection{Cosmological application of time asymmetric gravity}

 Let us now apply the time asymmetric extension of general relativity in a cosmological 
framework. In particular, we focus on a Friedmann-Robertson-Walker (FRW) spacetime metric 
of the form
\begin{equation}
ds^2 = -dt^2 + a(t)^2 \left(\frac{dr^2}{1-k r^2} + r^2 d\Omega^2\right) \,,
\end{equation}
where $a(t)$ is the scale factor, $k=-1,0,1$ for spatially open, flat or close geometry 
respectively, and with $d\Omega^2$ the two-dimensional sphere line element. 
Note that time-asymmetric extension of general relativity singles out a specific $3+1$ 
decomposition, selected by the constant mean curvature gauge condition, and moreover it 
introduces a dependence on the spatial slices volume, and thus the spacetime 
must be spatially compact. This is indeed the case in the above cosmological metric, 
where  $k$ refers to positive, negative or zero constant spatial curvature. In 
particular, all of these cases are consistent with a spatially compact topology, with 
$k=+1$ corresponding to spheres, $k=0$ to tori, while for $k=-1$  the infinite 
number of compact manifolds with constant negative curvature are classified by Thurston 
\cite{Thurston:1982zz}. Inserting the above metric in the total action $S+S_m$, with $S$ 
given by (\ref{newaction}) and $S_m$ the matter action, and performing the variation in 
the ADM formalism, we easily obtain the Friedmann equations as \cite{Cortes:2015ola}
\begin{equation}
H^2 + \frac{k}{a^2 }
=   \frac{8 \pi G   }{3  }\rho_m + f(V(a))^2
\label{FR100}
\end{equation}
\begin{equation}
\dot{H}-\frac{k}{a^2 }=  - 4 \pi G   (\rho_m+p_m)
 +af(V(a)) \frac{\partial f(V(a))}{\partial a},
 \label{FR200}
\end{equation}
where $H=\dot{a}/a$ is the Hubble parameter, $V(a)\propto a^3$ is the spatial volume, and 
$G$ is 
the gravitational constant. Additionally, we have considered the matter action $S_m$ to 
correspond to a perfect fluid with energy density $\rho_m$ and pressure $p_m$ 
respectively. We stress here that in action (\ref{newaction}) we do not include an 
explicit cosmological constant, since our goal is exactly to investigate 
whether the universe acceleration can arise solely from a general modification term 
$f(V)$ 
(which definitely in the specific case $f(V)=const.$ gives rise to an effective 
cosmological 
constant).

Defining for convenience $g(a)=\frac{a}{G}f(V(a))$, 
the 
above modified Friedmann equations become 
\begin{equation}
H^2 + \frac{k}{a^2 }
=   \frac{8 \pi G   }{3  }\rho_m + \frac{G^2 
g(a)^2}{a^2} 
\label{FR10}
\end{equation}
\begin{equation}
\dot{H}-\frac{k}{a^2 }=  - 4 \pi G   (\rho_m+p_m)
 +\frac{G^2 g(a)g^\prime(a)}{a}-\frac{G^2 
g(a)^2}{a^2},
 \label{FR20}
\end{equation}
and thus the modification is included in the arbitrary function $g(a)$. Furthermore, we 
can rewrite the Friedmann equations (\ref{FR10}),(\ref{FR20}) 
in the usual form 
\begin{subequations}
 \begin{eqnarray}
\label{Fr1}
H^2& =& \frac{\kappa^2}{3}\left(\rho_m + \rho_{DE} \right)   \\
\label{Fr2}
\dot{H}& =&-\frac{\kappa^2}{2}\left(\rho_m +p_m+\rho_{DE}+p_{DE}\right),
\end{eqnarray}
\end{subequations}
if we define the energy density and pressure of the effective dark energy sector as
 \begin{eqnarray}
\label{rhode}
&&\rho_{DE}\equiv \frac{3G}{8\pi}\frac{g(a)^2}{a^2} 
\\
&&p_{DE}  \equiv  -\frac{G}{8\pi}\left[\frac{g(a)^2}{a^2} +\frac{2 g(a)g^\prime(a)}{a}
\right],
\label{pde}
 \end{eqnarray}
i.e. attributing the dark energy sector to the new terms that time asymmetric 
gravity brings to the Friedmann equations. In this case, the dark energy 
equation-of-state parameter becomes:
\begin{eqnarray}
\label{wde}
w_{DE}\equiv 
\frac{p_{DE}}{\rho_{DE}}=-\frac{1}{3}\left[1+\frac{2ag^\prime(a)}{g(a)}\right].
\end{eqnarray}

In summary, the modified gravity at hand is determined by the arbitrary function 
$f(V(a))$. Hence, according to the choice of $f(V(a))$ one obtains distinct classes of 
cosmological models.

\section{Late-time cosmology}
\label{Dynanal}

In this section we are interested in investigating in detail the late-time cosmology of 
the time asymmetric extension of general relativity. Since the gravitational modification 
is determined by the function $f(V)$, we will choose two basic ansatzes, namely the 
power law and the exponential one. In particular, we will consider
\begin{itemize}
\item Model I:  $f(V)=g_1V^{m}$, which implies that the auxiliary function $g(a)$ becomes 
$g(a)=\frac{g_1}{G}\left(\frac{a}{a_0}\right)^p$, with $p=3m+1$,  with 
$g_1$ a constant and $p$ a parameter, and where $a_0$ is a constant which can be set to 1 
for convenience.

\item Model II:  $f(V)=g_2 e^{\lambda V}$, which implies that $g(a)=g_2\frac{a}{G} 
e^{\lambda a^3/a_0^3}$, with $g_2$ a constant, $\lambda$ a parameter, and with $a_0$ a 
constant which can be set to 1.
\end{itemize}

In order to study the cosmological behavior in a general way, independently of the
initial conditions and the specific universe evolution, we will apply the dynamical
systems method, which allows to extract the global features of a cosmological scenario
\cite{Perko,Ellis,Copeland:1997et,Ferreira:1997au,Chen:2008ft,Cotsakis:2013zha,
Giambo':2009cc,Xu:2012jf}. In this procedure, one first transforms the involved
cosmological equations into an autonomous system and then he extract its critical points. 
Hence, perturbing linearly around these critical points, and expressing the perturbations 
in terms of a perturbation matrix, allows to determine the type and stability of each 
critical point by examining the eigenvalues of this matrix.

\subsection{Model I:  $f(V)=g_1V^{m}$}
\label{ModIanalysis}

In the case where $f(V)=g_1V^{m}$, i.e. when $g(a)=\frac{g_1}{G}a^p$  (with $p=3m+1$), 
with $g_1$ a constant and $p$ a parameter, the Friedmann equations  
(\ref{FR10}),(\ref{FR20}) become 
\begin{equation}
H^2 + \frac{k}{a^2 }
=   \frac{8 \pi G   }{3  }\rho_m + g_1^2 a^{2p-2},
\label{FR1I}
\end{equation}
\begin{equation}
\dot{H}-\frac{k}{a^2 }=  - 4 \pi G   (\rho_m+p_m)
 +(p-1) g_1^2 a^{2p-2},
 \label{FR2I}
\end{equation}
and thus the effective dark energy (\ref{rhode}) and 
pressure (\ref{pde}) respectively  become 
\begin{eqnarray}
&&\rho_{DE}
=   \frac{3 g_1^2   }{ 8 \pi G } a^{2p-2}\\
&&p_{DE}
=   -\frac{ g_1^2   }{ 8 \pi G } a^{2p-2}(1+2p),
\end{eqnarray}
and hence (\ref{wde}) leads to
\begin{eqnarray}
\label{wdeI}
w_{DE} =-\frac{1}{3}(1+2p).
\end{eqnarray}
Additionally, we can define 
the ``total'' equation-of-state parameter 
\begin{equation}
w_{tot}\equiv -1-\frac{2\dot{H}}{3H^2}=\frac{8 \pi  G a^2 w_m  \rho_m-g_1^2 (2 p+1) a^{2 
p}+k}{8 \pi  G a^2 
\rho_m+3 g_1^2 
a^{2 p}-3k},
\end{equation}
and the deceleration parameter as
\begin{equation}
q\equiv -1-\frac{\dot{H}}{H^2}=\frac{1+3 w_{tot}}{2},
\label{deccelmodI}
\end{equation}
with $w_m\equiv p_m/\rho_m$ the matter equation of state. In the following we assume the 
usual energy conditions, which lead to $0\leq w_m \leq 1$. Finally, note that for 
$g_1=0$ we re-obtain standard general relativity.

\subsubsection{Zero or negative curvature}\label{Sect:3.1.1}

In the case $k=0,-1$ as auxiliary variables it proves convenient to use the various 
density parameters, namely 
\begin{align}
\Omega_k=-\frac{k}{a^2 H^2},\; \Omega_m=\frac{8 \pi G   \rho_m}{3  H^2},\; 
\Omega_{DE}=\frac{g_1^2 
a^{2p-2}}{H^2},
\label{auxmodIA}
\end{align}
and thus the first Friedmann equation (\ref{FR1I}) gives rise to the constraint
\begin{align}
\label{constr}
\Omega_k+\Omega_m+\Omega_{DE}=1.
\end{align}
Using the above auxiliary variables we can write the cosmological equations in the
autonomous form
\begin{subequations}
	\label{syst2}
	\begin{align}
	& \frac{d\Omega_k}{d\eta}=-\Omega_{k} \left[2 p \Omega_{DE}+(3 w_{m}+1)  
(\Omega_{DE}+\Omega_{k}-1)\right],\\
	&\frac{d\Omega_{DE}}{d \eta}=-\Omega_{DE} \left[2 p (\Omega_{DE}-1)+(3 w_{m}+1) 
(\Omega_{DE}+\Omega_{k}-1)\right],
	\end{align}
\end{subequations}
where we have used the constraint (\ref{constr}) in order to eliminate $\Omega_m$ and 
thus reduce the system to dimension two. In these equations, as usual, we define the 
logarithmic 
time $\eta=\ln a$. Hence, the above autonomous system is 
defined on the compact phase space
$\left\{(\Omega_k, \Omega_{DE}): \Omega_k\geq 0, \Omega_{DE}\geq 0, \Omega_k+ 
\Omega_{DE}\leq 1\right\}$ \footnote{The interest of defining compact phase spaces is that 
then the 
flow has well-defined past and future attractors, and this facilitates the drawing of 
global 
results for the cosmological scenario 
\cite{Perko,Ellis,Copeland:1997et,Ferreira:1997au,Chen:2008ft,
Cotsakis:2013zha,
Giambo':2009cc,Xu:2012jf}.}. Finally, using the auxiliary variables (\ref{auxmodIA}) we 
can express the deceleration parameter (\ref{deccelmodI}) as
 \begin{equation}
q=\frac{1}{2}\left[1+3w_m\Omega_m-(2p+1)\Omega_{DE}-\Omega_k\right].
\label{deccelmodIsol1}
\end{equation}
  
The scenario of Model I, namely $f(V)=g_1V^{m}$, with zero or negative curvature, admits 
three physical critical points, corresponding to expanding 
universe ($H>0$), which are displayed in Table \ref{Tab1} along with their existence 
conditions. In the same Table we include the eigenvalues of the involved perturbation 
matrix, and thus the corresponding stability conditions. Finally, for completeness, we 
also include the values of the deceleration parameter, calculated through 
(\ref{deccelmodIsol1}). Note that the solution associated to $P_3$ for $p\neq1$ is the 
power-law form $a(t)=\left[(1-p) \left(c_1+a_1 t\right)\right]^{\frac{1}{1-p}}$, while 
for 
$p=1$ it is just the de Sitter solution $a(t)=c_1 e^{g_1 t}$, with $c_1$ and $a_1$ 
integration constants.

In summary, the scenario at hand admits two stable late-time critical points, namely 
$P_2$ 
 for $p<0$ and $P_3$  for $p>0$.
 \begin{table*}
	\resizebox{\columnwidth}{!}
	{
\!\!\!\!\!\!\!\!\!\!\!\!\!\begin{tabular}{ccccccc}
	\hline  
	C.P. & $\Omega_k$& \!\!\!$\Omega_{DE}$ \!\!\!&$q$ &\!\!\! Existence &\!\!\! 
Eigenvalues & \!\!\!Stability \\ 
	\hline  
	$P_1$ & $0$ & \!\!\!$0$ & \!\!\!$\frac{3w_m+1}{2}$  & \!\!\!always  & 
\!\!\!{\small{$3 w_m+1,2 p+3 w_m+1$}} &\!\!\! {\small{saddle for  $p<-\frac{3w_m+1}{2}$}} 
\\
	      &&&&&&  \!\!\! {\small{unstable for $0\leq w_{m}\leq 1, 
p>-\frac{3w_m+1}{2}$}}\\ 
	      	\hline  
	$P_2$ & $1$ &\!\!\! $0$ & \!\!\!$0$ & \!\!\!always\!\!\!  &{\small{ $2 p,-3 
w_m-1$ 
}}& \!\!\! {\small{stable for $p<0$ }}
\\
	      	&&&&&&
	      \!\!\!	 {\small{ saddle for $ p>0$}} \\ 
	\hline  
	$P_3$& $0$ & \!\!\!$1$ & \!\!\!$-p$  &  \!\!\! always &  \!\!\!   {\small{ $-2 
p,-2 p-3 w_m-1$ }} \!\!\! &{\small{ unstable for $p<-\frac{3w_m+1}{2}$ }}\\
	 	&&&&&& \!\!\! {\small{saddle for  $-\frac{3w_m+1}{2}<p<0$}}  \\
	 	&&&&&& \!\!\!{\small{stable for $p>0$}}\\
	\hline 
\end{tabular} 
}\caption{\label{Tab1} 
The physical critical points of the system   \eqref{syst2} of time asymmetric cosmology 
of Model I:  $f(V)=g_1V^{m}$, with zero or negative curvature, and their existence and 
stability conditions. We have assumed $0\leq w_{m}\leq 1$.}
\end{table*}

\subsubsection{Positive curvature}

In the case $k=+1$, that is for positive curvature, it is not guaranteed that the Hubble 
parameter does not change sign during the evolution. This implies that the 
$H$-normalization that we used in the previous open and flat case is not a good choice 
for creating compact variables, since when $H$ crosses zero the dynamical variables would
diverge, and moreover when $H$ change sign our ``time'' variable $\eta=\ln a$ would 
change flow.  Thus, in the present $k=+1$ case, it is consistent to introduce the 
auxiliary variables (similarly to the variables introduced in section VI of
\cite{Coley:2003mj},  and in sections 3.3 and 5.3 of \cite{Leon:2009ce}) as:
\begin{align}
\Theta_k=\frac{1}{a^2 \mathcal{D}^2},\; Q_0=\frac{H}{\mathcal{D}},\; \Theta_m=\frac{8 \pi 
G 
\rho_m}{3 
 \mathcal{D}^2},\; \Theta_{DE}=\frac{g_1^2 a^{2(p-1)}}{\mathcal{D}^2},
 \label{auxmodIB}
\end{align}
where $\mathcal{D}=\sqrt{H^2+ a^{-2}}$, and which are finite even if $H$ crosses 
zero. Therefore, the first Friedmann equation (\ref{FR1I}) leads to the constraint
\begin{equation}
\label{constr2}
\Theta_m+\Theta_{DE}=1.
\end{equation}
Additionally, from the definition of $D$ it follows 
\begin{equation}
\label{constr3}
\Theta_k+Q_0^2=1,
\end{equation}
while the curvature parameter is expressed as $$\Omega_k\equiv \frac{1}{a^2 
H^2}=\frac{1-Q_0^2}{Q_0^2}.$$ 

Using the above auxiliary variables we can re-write the cosmological equations as
\begin{subequations}
	\label{syst4}
	\begin{align}
	& \frac{d Q_0}{d \tau}=\frac{1}{2} \left(1-Q_0^2\right) \left[2 p \Theta_{DE}+(3 
w_m +1)(\Theta_{
DE}-1)\right],\\
	&\frac{d \Theta_{DE}}{d\tau}= -Q_0 (2 p+3 w_m+1)(\Theta_{DE}-1) \Theta_{DE},
	\end{align}
\end{subequations}
where we have used the constraints  \eqref{constr2} and \eqref{constr3} in order to 
eliminate $\Theta_m$ and $\Theta_k$ and therefore reduce the system to dimension two.
In the above dynamical system, we have introduced the consistent ``time'' variable 
$\tau$ through $d\tau=Ddt$, which indeed satisfies the necessary requirement that 
it is monotonic even if $H$ change sign.  The 
above autonomous system is 
defined on the compact phase space $\left\{(Q_0, \Theta_{DE}): -1\leq 
Q_0\leq 1, 0\leq 
\Theta_{DE}\leq 1\right\}$.
Finally, using the auxiliary variables (\ref{auxmodIB}) we 
can express the deceleration parameter (\ref{deccelmodI}) as
 \begin{equation}
q=\frac{1}{2 Q_0^2}\left[1+3w_m 
(1-\Theta_{DE}) -(2p+1)\Theta_{DE}\right].
\label{deccelmodIsol2}
\end{equation}

As we can observe the system \eqref{syst4} is symmetric under the transformation 
\begin{equation}
\label{discrete_symm}
(\tau, Q_0, \Theta_{DE})\rightarrow (-\tau, -Q_0, \Theta_{DE}).
\end{equation}
Thus, it is sufficient to discuss the behavior in one part of the phase space, that is in
$\tau\geq 0, 
Q_0\geq 0, \Theta_{DE} \geq 0$, and then  obtain the dynamics on the other part from 
\eqref{discrete_symm}. For example, if a point with coordinates $(Q_0^*,\Theta_{DE}^*), 
Q_0^*>0, \Theta_{
DE}^*>0$ is a future attractor as $\tau\rightarrow +\infty$, then its partner point 
$(-Q_0^*,\Theta_{DE}^*)$ via \eqref{discrete_symm} is a past attractor as 
$\tau\rightarrow 
-\infty$, and vice versa. 
Furthermore, we mention that the function 
\begin{equation}
M=\frac{1-\Theta_{DE}}{1-Q_0^2}, \quad \frac{d M}{d\tau}=- (3 w_m+1) Q_0 M,
\end{equation} 
is a monotonic function in the regions $Q_0<0$ and $Q_0>0$ for $\Theta_{DE}\neq 1$. The 
points having $Q_0>0$ correspond to expansion, while those having $Q_0<0$ correspond to  
contraction. As we will see in the following, the system \eqref{syst4} admits a fixed 
point with $Q_0=0$ if we assume $p\geq 0, 0\leq w_m\leq 1$. However, since for $p<0, 
0\leq 
w_m\leq 1$ there are not equilibrium points with $Q_0=0$, it follows that $M$ acts as a 
monotonic function in the interior of the phase space. As a consequence, for $p<0$ there 
can be no 
periodic orbits in the interior of the phase space and global results can be drawn 
\cite{Coley:2003mj}. 
Additionally, from the definition of $M$ it follows that either $Q_0^2\rightarrow 1$ 
or $\Theta_{DE}\rightarrow 1$ asymptotically.  

Note that the system  \eqref{syst4} allows for an easy analytical elaboration, leading to 
 \begin{equation}
 \label{exact}
 Q_0^2(a) =1-\frac{c_2 a^{3 w_{m}+1}}{a^{2 p+3 w_{m}+1}+e^{c_1}}, \ \ \ \ \ \
 \Theta_{DE}(a)=1-\frac{e^{c_1}}{a^{2 p+3 w_{m}+1}+e^{c_1}},
 \end{equation}
 with $c_1$ and $a_1$ integration constants.

The scenario at hand admits five physical critical points which are displayed in Table 
\ref{Tab2} along with their existence conditions. In the same Table we include the 
eigenvalues of the corresponding perturbation matrix, and the resulting stability 
conditions. Finally, we also include the values of the deceleration parameter, calculated 
through (\ref{deccelmodIsol2}). Note that the points $P_4$ and $ P_5$ have the time 
reversal behavior of $P_6$ and $ P_7$ respectively, due to the symmetry 
\eqref{discrete_symm}. Additionally, there exist orbits connecting $P_4$ and $
P_5$ with  $P_6$ and $ P_7$, which implies that $Q_0$ can indeed become zero, i.e. $H=0$, 
during the evolution (recall that the points having $Q_0>0$ are expanding while those 
having $Q_0<0$ are contracting). Lastly, the system admits a static solution, 
namely $P_8$, which always behaves as a saddle point.

In summary, the scenario of Model I, namely $f(V)=g_1V^{m}$, with positive 
curvature, admits three stable late-time critical points, namely 
the expanding solution $P_5$ for $p>0$, the contracting solution $P_6$ for $p > -(3w_m 
+1)/2$ and the contracting solution $P_7$ for $p < -(3w_m+1)/2$.
\begin{table*}
	\resizebox{\columnwidth}{!}
	{
\!\!\!\!\!\!\!\!\!\!\!\!\!\begin{tabular}{ccccccc}
			\hline  
C.P. & $Q_0$& \!\!\!$\Theta_{DE}$ \!\!\!&$q$ &\!\!\! Existence 
&\!\!\! 
Eigenvalues & \!\!\!Stability\\ 
			\hline  
$P_4$ & $1$ &\!\!\! $0$ & \!\!\!$\frac{3w_m+1}{2}$   &\!\!\! always  & \!\!\!{\small{  $3 
w_m+1,2 p+3 w_m+1$}} & \!\!\!{\small{ saddle for 
$p<-\frac{3w_m+1}{2}$ }}\\
			&&&&&&   \!\!\!{\small{ unstable for $p>-\frac{3w_m+1}{2}$}}   \\ 
			\hline  
			$P_5$ & $1$ &\!\!\! $1$ & \!\!\!$-p$ & \!\!\!always  & 
\!\!\!{\small{$-2 p,-2 p-3 w_m-1$}} &  \!\!\!{\small{  unstable for 
$p<-\frac{3w_m+1}{2}$ }} \\
			&&&&&&  \!\!\!{\small{saddle for  $-\frac{3w_m+1}{2}<p<0$}}  \\
			&&&&&& \!\!\!{\small{  stable for $p>0$}}\\
			\hline 
$P_6$ & $-1$ &\!\!\! $0$ & \!\!\!$\frac{3w_m+1}{2}$   &\!\!\! always  & \!\!\!{\small{  
$-(3 
		w_m+1),-(2 p+3 w_m+1)$}} & \!\!\!{\small{ saddle for 
		$p<-\frac{3w_m+1}{2}$ }}\\
&&&&&&   \!\!\!{\small{ stable for $p>-\frac{3w_m+1}{2}$}}   \\ 
\hline 	
	$P_7$ & $-1$ &\!\!\! $1$ & \!\!\!$-p$ & \!\!\!always  & 
	\!\!\!{\small{$2 p,2 p+3 w_m+1$}} &  \!\!\!{\small{  stable for 
			$p<-\frac{3w_m+1}{2}$ }} \\
	&&&&&&  \!\!\!{\small{saddle for  $-\frac{3w_m+1}{2}<p<0$}}  \\
	&&&&&& \!\!\!{\small{  unstable for $p>0$}}\\
	\hline 
	$P_8$ & $0$ & $\frac{3 w_m+1}{2 p +3 w_m+1}$ & undefined & $p\geq 0$ & 
$-\sqrt{p(1+3w_m)}, \sqrt{p(
1+3w_m)}$ & saddle 	\\
	\hline			
			\end{tabular} 
		}
\caption{\label{Tab2}
The physical critical points of the system \eqref{syst4} of time asymmetric cosmology 
of Model I: $f(V)=g_1V^{m}$, with positive curvature, and their existence and 
stability conditions.  } 
\end{table*}

We close this paragraph with some comments on the auxiliary variables choice. The 
advantage of using the variable $Q_0$ versus using the variable $\Omega_k$, used
in paragraph \ref{Sect:3.1.1}, is that for closed models the variable $\Omega_k$ would 
not keep track of the $H$-sign changes, due to the quadratic dependence on $Q_0$, however 
these changes may have important cosmological consequences. As we saw, choosing $Q_0$ 
instead of $\Omega_k$ allows us to to obtain novel features, such as expanding solutions, 
contracting partners, transition from contracting to expanding cosmologies and vice 
versa, as well as static solutions (for instance a static solution, where 
$H=0$, would obviously not be seen using $H$-normalization). These differences, arising 
from the possible $H$-sign change in closed models, forbids a unified description of all 
cases. Such a difference between closed and open/flat geometries, and the implied 
necessary different normalization, was first observed in 
\cite{Coley:2000yc,Goliath:1998na,Coley:2003mj} despite the fact that closed FRW had been 
previously studied in   \cite{Halliwell:1986ja,Ellis,vandenHoogen:1999qq}.

\subsection{Model II:   $f(V)=g_2 e^{\lambda V}$}
\label{ModIIanalysis}

In the case where $f(V)=g_2 e^{\lambda V}$, i.e. when $g(a)=g_2\frac{a}{G} 
e^{\lambda a^3}$, with $g_2$ a constant and $\lambda$ a parameter, the Friedmann 
equations  
(\ref{FR10}),(\ref{FR20}) become 
\begin{equation}
H^2 + \frac{k}{a^2 }
= \frac{8 \pi  G \rho_m}{3} + g_2^2 e^{2 \lambda a^3   }
\label{FR1II}
\end{equation}
\begin{equation}
\dot{H}-\frac{k}{a^2 }=  - 4 \pi G   (\rho_m+p_m)
+3 a^3 g_2^2 \lambda  e^{2 \lambda a^3  },
\label{FR2II}
\end{equation}
 and thus
 \begin{eqnarray}
 \label{wdeII}
 w_{DE} =-1-2 \lambda a^3.
 \end{eqnarray} 
 Additionally, the ``total'' equation-of-state parameter reads 
\begin{equation}
w_{tot}\equiv -1-\frac{2\dot{H}}{3H^2}=\frac{8 \pi  G a^2 w_m  \rho_m- 3 g_2^2 a^2  e^{2 
\lambda a^3  }(1+2 \lambda a^3) +k}{8 \pi  G a^2 
\rho_m+3 g_2^2 a^2  e^{2 
\lambda a^3  }  -3k},
\end{equation}
while the deceleration parameter writes as
\begin{equation}
q=\frac{1+3 w_{tot}}{2}.
\label{deccelmodII}
\end{equation}
 Finally, note that for $g_2=0$ we 
re-obtain standard general relativity.

 \subsubsection{Zero or negative curvature}

In the case $k=0,-1$, we introduce the density parameters compact auxiliary variables 
\begin{align}
\Omega_k=-\frac{k}{a^2 H^2},\; \Omega_m=\frac{8 \pi G   \rho_m}{3  H^2},\; 
\Omega_{DE}=\frac{g_2^2 
e^{2 a^3 \lambda }}{H^2},
\label{auxmodII}
\end{align}
and thus the first Friedmann equation (\ref{FR1II}) gives rise to the constraint
\begin{align}
\label{constrbbb}
\Omega_k+\Omega_m+\Omega_{DE}=1.
\end{align}
In order to be able to close the system we need one more auxiliary parameter. Since the 
corresponding choice proves to be different according to the sign of $\lambda$, we will 
examine the two cases separately.\\
\begin{itemize}

\item  $\lambda>0$

In this case we define the additional auxiliary variable
 \begin{align}
\label{auxmodIIT}
 T=\frac{\lambda a^3}{1+\lambda a^3}.
\end{align}
 Since by construction  $0< T< 1$ (since $\lambda>0$), we can define  
\begin{equation}
\frac{d\bar{\eta}}{dt}=H(1-T)^{-1},
\end{equation}
which implies $\bar{\eta}=\frac{1}{3} \lambda  a(t)^3+\ln [a(t)]$ (modulo an additive 
constant), and thus $\bar{\eta}\rightarrow -\infty$ as $a\rightarrow 0$  
and $\bar{\eta}\rightarrow  \infty$ as $a\rightarrow \infty$. Hence, using the auxiliary 
variables (\ref{auxmodII}) and (\ref{auxmodIIT}) we can re-write the cosmological 
equations in their autonomous form, namely 
\begin{subequations}
\label{systII2}
\begin{align}
& \frac{d T}{d\bar{\eta}}=3 T(1-T)^2,
\label{eqT}\\
& \frac{d\Omega_k}{d\bar{\eta}}=3 (T-1) (w_{m}+1) \Omega_{k} 
(\Omega_{DE}+\Omega_{k}-1)-2 \Omega_
{k} \left[T (3 \Omega_{DE}+\Omega_{k}-1)-\Omega_{k}+1\right],\\
&\frac{d\Omega_{DE}}{d\bar{\eta}}=3 (T-1) (w_{m}+1) \Omega_{DE} 
(\Omega_{DE}+\Omega_{k}-1)+2 \Omega_{DE} \left[\Omega_{k}-T (3 
\Omega_{DE}+\Omega_{k}-3)\right],
\end{align}
\end{subequations}
where we have used the constraint (\ref{constrbbb}) in order to eliminate $\Omega_m$. 
Clearly, the above system is defined on the 
$\left\{(T,\Omega_k, \Omega_{DE}): 0\leq T\leq 1, \Omega_k\geq 0, \Omega_{DE}\geq 0, 
\Omega_k+ \Omega_{DE}\leq 1\right\}$ part of the phase space, where we have included the 
two 
boundaries $T=0$ and $T=1$. Furthermore, note that the 
invariant subset boundary $T=1$ corresponds to the asymptotic future, while the invariant 
subset boundary $T=0$ is associated asymptotically to the (classical) initial state. 
Therefore, in this formalism all the fixed points are located at $T=0$ and $T=1$ 
\cite{Alho:2015cza}. Lastly, using the auxiliary variables (\ref{auxmodII}), 
(\ref{auxmodIIT}) we 
can express the deceleration parameter (\ref{deccelmodII}) as
 \begin{equation} 
q=\frac{1}{2}\left[1+3w_m\Omega_m-3\left(\frac{1+T}{1-T}\right)\Omega_{DE}-\Omega_k\right]
.
\label{deccelmodIIsol1}
\end{equation}
Thus, for the critical points having $T=0$, the expression \eqref{deccelmodIIsol1} is 
well-defined and gives $q\left|_{T=0}\right. =  
\left[1+3w_m\Omega_m-3\Omega_{DE}-\Omega_k\right]/2$, while
for the critical points having $T=1, \Omega_{DE}>0$ we obtain
$q\rightarrow -\infty$ as  $T\rightarrow 1^-$ since $\Omega_m, \Omega_{DE}$ and 
$\Omega_k$ are bounded. 
On the other hand, for the critical points having $T=1,\Omega_{DE}=0$, $q$ is 
arbitrary. 

$\ \ \,$The scenario at hand admits five physical critical points, and one curve of 
critical points (namely $Q_5$), which are summarized in Table \ref{Tab3} along with their 
existence conditions. In the same Table we include the eigenvalues of the corresponding 
perturbation matrix, and the resulting stability conditions. Finally, we also include the 
values of the deceleration parameter, calculated through (\ref{deccelmodIIsol1}). We 
mention 
that for the three nonhyperbolic critical points, the linear analysis is not adequate to 
determine their stability, and therefore the stability conditions have been extracted 
applying the center manifold method \cite{wiggins}. The corresponding investigation is 
performed in Appendix \ref{App1}.
\begin{table*}
\resizebox{\columnwidth}{!}
{
\!\!\!\!\!\!\!\!\!\!\!\!
\begin{tabular}{cccccccc}
\hline  
	Label & $\Omega_k$& $\Omega_{DE}$ & $T$ & $q$ & Existence & 
Eigenvalues 
& Stability \\ 
\hline  
	$Q_1$ & $0$ & $0$ & $0$ & $\frac{3w_m+1}{2}$ & always & {\small{$3, 
3(1+w_m), 1+3 w_m$}} & 
unstable 
\\ 
\hline
	$Q_2$ & $1$ & $0$ & $0$  &$0$& always & {\small{$3, -3(1+w_m), 2$ }}& 
saddle \\ 
\hline 
$Q_3$ & $0$ & $1$ & $0$  &{\small{ $-1$}}  & always 
& {\small{$3, -3(1+w_m), -2$ }}& saddle \\ 
\hline
$Q_4$ & $0$ & $0$ & $1$ &{\small{arbitrary}} & always & $6, 0, 0$ & 
{\small{ nonhyperbolic,}}
\\
&&&&&&&  {\small{ behaves as unstable}} \\ 
\hline
	$Q_5$ & $\Omega_{k c}$ & $0$ & $1$ & {\small{arbitrary}}  & 
$\Omega_{k c}\in (0,1]$ & 
$6, 
0, 0$ & {\small{nonhyperbolic,}}	\\
			&&&&&&&  {\small{ behaves as unstable}}\\ 
  			\hline
  			$Q_6$ & $0$ & $1$ & $1$  & $-\infty$  & always & $-6, -6, 
0$ & {\small{nonhyperbolic,}}
  			\\
			&&&&&&&   {\small{ behaves as stable}}\\ 
  			\hline
  		\end{tabular} }
 \caption{\label{Tab3} The physical critical points of the system   \eqref{systII2} of 
time asymmetric cosmology of Model II:  $f(V)=g_2 e^{\lambda V}$, with zero or negative 
curvature and $\lambda>0$, and their existence and stability conditions. We assume $0\leq 
w_m\leq 
1$.}
  \end{table*}

\item  $\lambda<0$

In this case we define the additional auxiliary variable
 \begin{align}
\label{auxmodIITb}
  {T_{1}}=-\frac{\lambda a^3}{1-\lambda a^3}.
\end{align}
 Since    $0< {T_{1}}< 1$ (since $\lambda<0$), we can define  
\begin{equation}
  \frac{d\check{\eta}}{dt}=H(1-{T_{1}})^{-1},
\end{equation}
which implies $\check{\eta}=-\frac{1}{3} \lambda  a(t)^3+\ln [a(t)]$ (modulo an additive 
constant), and thus $\check{\eta}\rightarrow -\infty$ as $a\rightarrow 0$ 
and $\check{\eta}\rightarrow  \infty$ as $a\rightarrow \infty$. Hence, using the 
auxiliary variables (\ref{auxmodII}) and (\ref{auxmodIITb}) we can re-write the 
cosmological equations in autonomous form
as
\begin{subequations}
  	\label{systII2b}
  	\begin{align}
&\frac{d {T_{1}}}{d\check{\eta}}=3{T_{1}}(1-{T_{1}})^2,\label{II2b.1}\\
&\frac{d \Omega_k}{d\check{\eta}}=3\Omega_{k} \left[ ({T_{1}}-1) w_m-1\right] 
(\Omega_{DE}+\Omega_{k}-1)+\Omega_{k} \left[{T_{1}} (9 
\Omega_{DE}-1)-2\right]+({T_{1}}+2) 
\Omega_{k}^2,\\
&\frac{d \Omega_{DE}}{d\check{\eta}}=\Omega_{DE} \left\{3 (\Omega_{DE}-1) \left[{T_{1}} 
(w_m+3)-w_m-1\right]+({T_{1}}-1) (3 w_m+1) \Omega_{k}\right\},
\end{align}
  \end{subequations}
where we have used the constraint (\ref{constrbbb}) in order to eliminate $\Omega_m$. 
Clearly, the above system is defined on the 
$ \left\{({T_{1}},\Omega_k, \Omega_{DE}): 0\leq {T_{1}}\leq 1, \Omega_k\geq 0, 
\Omega_{DE}\geq 
0, 
  \Omega_k+ \Omega_{DE}\leq 1\right\}$ part of the phase space, where we have attached 
the 
two 
boundaries ${T_{1}}=0$ and ${T_{1}}=1$. Finally, note that 
in terms of the auxiliary variables (\ref{auxmodII}),(\ref{auxmodIITb}) the deceleration 
parameter (\ref{deccelmodII}) is given by
\begin{equation}
q=\frac{1}{2}\left[1+3w_m\Omega_m-3\left(\frac{1-3{T_{1}}}{1-{T_{1}}}\right)\Omega_{
DE} -\Omega_
k\right].
\label{deccelmodIIsol1b}
\end{equation}
Thus, for the critical points having ${T_{1}}=0$, the expression 
\eqref{deccelmodIIsol1b} is well-defined 
and gives 
$q\left|_{{T_{1}}=0}\right. = 
\left[1+3w_m\Omega_m-3\Omega_{DE}-\Omega_k\right]/2$, while  for the critical 
points having ${T_{1}}=1, \Omega_{DE}>0$  it follows that $q\rightarrow 
+\infty$ as  ${T_{1}}\rightarrow 1^-$ since $\Omega_m, \Omega_{DE}$ and $\Omega_k$ are 
bounded. On the other hand, for the critical points having ${T_{1}}=1,\Omega_{DE}=0$, 
$q$ is arbitrary.
  
$\ \ \,$The scenario at hand admits five physical critical points, and one curve of 
critical points (namely $Q_{11}$), which are summarized in Table \ref{Tab3b} along with 
their existence conditions. In the same Table we present the eigenvalues of the 
involved perturbation matrix and the corresponding stability conditions. 
Finally, we also include the 
values of the deceleration parameter, calculated through (\ref{deccelmodIIsol1b}). 
Concerning the three nonhyperbolic critical points the stability conditions have been 
extracted applying the center manifold method \cite{wiggins}. The corresponding 
investigation is performed in Appendix \ref{App2}.
\begin{table*}
    \resizebox{\columnwidth}{!}
{
\!\!\!\!\!\!\!\!\!\!\!
    	\begin{tabular}{cccccccc}
\hline  
	Label & $\Omega_k$& $\Omega_{DE}$ & ${T_{1}}$ &$q$ & Existence & Eigenvalues 
	& \!\!\!\!\!\!\!\!\!\!\!\!\!\!\!\!\!\!Stability \\ 
	\hline  
$Q_7$ & $0$ & $0$ & $0$  & $\frac{3w_m+1}{2}$ & always & {\small{ $3, 3(1+w_m), 1+3 w_m$ 
}} & \!\!\!\!\!\!\!\!\!\!\!\!\!\!\!\!\!\!unstable 
\\ 
\hline
  $Q_8$ & $1$ & $0$ & $0$ &$0$ & always & {\small{ $3, -3(1+w_m), 2$}} & 
\!\!\!\!\!\!\!\!\!\!\!\!\!\!\!\!\!\!saddle \\ 
  	\hline 
 	$Q_9$ & $0$ & $1$ & $0$ & {\small{ $-1$}}  & always &  {\small{ $3, 
-3(1+w_m), -2$}} & \!\!\!\!\!\!\!\!\!\!\!\!\!\!\!\!\!\!saddle \\ 
\hline
$Q_{10}$ & $0$ & $0$ & $1$& {\small{arbitrary}}   & always & $-6, 0, 0$ &   
\!\!\!\!\!\!\!\!\!\!\!\!\!\!\!\!\!\!{\small{ nonhyperbolic,}}
 	\\
	&&&&&&&  \!\!\!\!\!\!\!\!\!\!\!\! \!\!\!\!\!\!{\small{ behaves as stable }}\\ 
	\hline
 	$Q_{11}$ & $\Omega_{k c}$ & $0$ & $1$  &  {\small{arbitrary}}  & 
$\Omega_{kc}\in (0,1]$ & 
$-6, 
 	0, 0$ & \!\!\!\!\!\! \!\!\!\!\!\!\!\!\!\!\!\!{\small{nonhyperbolic,}}
 		\\
	&&&&&&& \!\!\!\!\!\!\!\!\!\!\!\! \!\!\!\!\!\!{\small{ behaves as saddle for 
$\Omega_{k c}\neq 1$.  
}}\\ 
	&&&&&&& \!\!\!\!\!\!\!\!\!\!\!\! \!\!\!\!\!\!{\small{ stable for $\Omega_{k c}= 
1$.  }}\\
	\hline
$Q_{12}$ & $0$ & $1$ & $1$  &{\small{ $+\infty$}}  & always & $6, 6, 0$ & 
 {\small\!\!\!\!\!\!\!\!\!\!\!\!\!\!\!\!\!\!{nonhyperbolic,}}
	\\
&&&&&&&  \!\!\!\!\!\!\!\!\!\!\!\!\!\!\!\!\!\! {\small{ behaves as unstable }}  \\
  \hline
    	\end{tabular} }
    	\caption{\label{Tab3b}
    	The physical critical points of the system   \eqref{systII2b} of 
time asymmetric cosmology of Model II:  $f(V)=g_2 e^{\lambda V}$, with zero or negative 
curvature and $\lambda<0$, and their existence and stability conditions. We assume $0\leq 
w_m\leq 
1$. }
    \end{table*}

  \end{itemize}
    
In summary, the scenario  of Model II, namely  $f(V)=g_2 e^{\lambda V}$, with zero or 
negative curvature admits the following stable late-time solutions: For $\lambda>0$ the 
expanding solution $Q_6$ (nonhyperbolic, but with stable center manifold). For 
$\lambda<0$ the line of expanding solutions $Q_{10}$ (nonhyperbolic, but with stable 
center manifold) and a point on the line  $Q_{11}$ with $\Omega_{k c}=1$  
(nonhyperbolic, but with stable center manifold)  which represent a curvature dominated 
solution. 
  
\subsubsection{Positive curvature}

In this case we introduce the density parameters compact auxiliary variables (similarly 
to the variables introduced in section VI of \cite{Coley:2003mj}, and in sections 3.3 
and 5.3 of \cite{Leon:2009ce}):
 \begin{align}\label{varsIIa}
 \Theta_k=\frac{1}{a^2 \mathcal{D}^2},\; Q_0=\frac{H}{\mathcal{D}},\;  \Theta_m=\frac{8 
\pi 
G   
\rho_m}{3  \mathcal{D}^2},\; \Theta_{DE}=\frac{g_2^2 e^{2 a^3 \lambda 
}}{\mathcal{D}^2},
 \end{align}
 with $\mathcal{D}=\sqrt{H^2+ a^{-2}}$,
and thus the first Friedmann equation (\ref{FR1II}) gives rise to the constraint
\begin{align}
\label{constrII2}
\Theta_m+\Theta_{DE}=1,
\end{align}
and as before we have the restriction \begin{equation}
\label{constrII3}
\Theta_k+Q_0^2=1.
\end{equation}
In order to be able to close the system we need one extra auxiliary parameter. Since the 
corresponding choice is different for different signs of $\lambda$, we will 
examine the two cases separately.\\

\begin{itemize}

\item $\lambda>0$

In this case we define the additional auxiliary variable
 \begin{align}
\label{auxmodIITpositive}
 T=\frac{\lambda a^3}{1+\lambda a^3}.
\end{align}
 Since by construction $\mathcal{D}\geq 0$ and   $0< T< 1$ (since 
$\lambda>0$), we can define  
\begin{equation}
\frac{d\bar{\tau}}{dt}=\mathcal{D}(1-T)^{-1},
\end{equation}
which implies $\bar{\tau}=\int_1^t \frac{\left(\lambda  a(\zeta)^3+1\right) 
\sqrt{a'(\zeta)^2+1}}{a(\zeta)} \, d\zeta$ (modulo an additive 
constant). Since $\mathcal{D}(1-T)^{-1}>0$, $\bar{\tau}$ is a monotonic function of $t$. 
Hence, 
using the auxiliary 
variables (\ref{varsIIa}) and (\ref{auxmodIITpositive}) we can express the cosmological 
equations as
\begin{subequations}
	\label{systII4}
	\begin{align}
	&\frac{d T}{d\bar{\tau}}=3 Q_0 T(1-T)^2,\\
	&\frac{d Q_0}{d\bar{\tau}}=\frac{1}{2} \left(1-Q_0^2\right) \left[-3 (T-1) 
(w_m+1) 
(\Theta_{DE}-1)
+T (6\Theta_{DE}-2)+2\right],\\
	&\frac{d\Theta_{DE}}{d\bar{\tau}}=3 Q_0 (\Theta_{DE}-1) 
\Theta_{DE} \left[T (w_m-1)-
w_m-1\right],
	\end{align}
\end{subequations}
where we have used the constraint (\ref{constrII2}) in order to eliminate 
$\Theta_m$ and \eqref{constrII3} to eliminate $\Theta_k$. 
The above system is defined on the $\Big\{(T,Q_0, \Theta_{DE}): 0\leq T \leq 1,  -1\leq 
Q_0\leq 1, 0\leq \Theta_{DE}\leq 1\Big\}$ part of the phase space, 
where we have included the two boundaries $T=0$ and $T=1$. 
For $Q_0>0$, i.e. for expanding cosmologies, the invariant subset boundary $T=1$ 
corresponds to the asymptotic future, while the invariant subset boundary $T=0$ is 
associated asymptotically to the (classical) initial state. However, for contracting 
models ($Q_0<0$) the roles of the invariant sets $T=1$ and $T=0$ are reversed in time. 
This arises from the fact that the function   
\begin{equation}
N=\frac{T}{1-T}, \quad N'=3 (1-T) Q_0 N,
\end{equation}  
is a monotonic function in the region $Q_0<0, 0<T<1$, where $N$ is monotonically 
increasing, and in the region $Q_0>0, 0<T<1$, where $N$ is monotonically decreasing, and 
thus it follows that either $T\rightarrow 1$ or $T\rightarrow 0$ asymptotically. 
Hence, in this formalism all the fixed points are located at $T=0$ and $T=1$ 
\cite{Alho:2015cza}. 

Furthermore, the system \eqref{systII4}   is invariant under the symmetry 
\begin{equation}
\label{discrete_symm_II}
\left(\bar{\tau}, T, Q_0,  \Theta_{DE}\right)\rightarrow \left(-\bar{\tau}, T, -Q_0,  
\Theta_{DE}\right).
\end{equation}
Thus, it is sufficient to discuss the behavior in one part of the phase space, for 
instance in $\bar{\tau}\geq 0, T\geq 0, Q_0\geq 0, \Theta_{DE} \geq 0$, while the 
dynamics on the other part is being obtained from \eqref{discrete_symm_II}. For example, 
if a point with coordinates $(T^*, 
Q_0^*,\Theta_{DE}^*),
 Q_0^*>0, \Theta_{DE}^*>0$ is a future attractor as $\bar{\tau}\rightarrow +\infty$, 
then, 
its partner point $(T^*,-Q_0^*,\Theta_{DE}^*)$ via \eqref{discrete_symm_II} is a past 
attractor as $\bar{\tau}\rightarrow -\infty$, and vice versa. 

Finally, using the auxiliary variables (\ref{varsIIa}) and (\ref{auxmodIITpositive}) we 
can express the deceleration parameter (\ref{deccelmodII}) as
 \begin{equation}
  q= \frac{1}{2Q_0^2}\left[1+3w_m 
(1-\Theta_{DE}) -3\left(\frac{1+T}{1-T}\right)\Theta_{DE}\right].
\label{deccelmodIIsol2b}
\end{equation}
 Thus, for the critical points having $T=0$, the expression \eqref{deccelmodIIsol2b} is 
well-defined and gives 
$ q\left|_{T=0}\right. = \frac{1}{2Q_0^2}\left[1+3w_m 
 (1-\Theta_{DE}) -3 \Theta_{DE}\right]$, whereas for the critical 
points having $T=1, \Theta_{DE}\neq 0$, it is implied that $q\rightarrow -\infty$ 
as  
$T\rightarrow 1^-$. On the other hand, for the critical points having 
$T=1,\Theta_{DE}=0$, $q$ is arbitrary.   

Furthermore, we note that the function 
\begin{equation}
M=\frac{1-\Theta_{DE}}{1-Q_0^2}, \quad M'=- (3 w_m+1)(1-T) Q_0 M,
\end{equation} 
is a monotonic function in the regions $Q_0<0, 0<T<1$ and $Q_0>0, 0<T<1$ for 
$\Theta_{DE}\neq 1$. 
The points having $Q_0>0$ are expanding, and those having $Q_0<0$ are contracting.  Thus, 
from the definition of $M$ it follows that either $Q_0^2\rightarrow 1$ or 
$\Theta_{DE}\rightarrow 1$ asymptotically. 
\\
The scenario at hand admits four physical critical points, and one curve of critical 
points (namely $Q_{16}$, which is the straight line joining the points $(0,0,1)$ 
and $(1,0,1)$, with the left point not included) corresponding to expanding 
cosmologies, which are summarized in Table \ref{Tab4} along with their existence 
conditions. In the same Table we include the eigenvalues of the 
involved perturbation matrix, and the corresponding stability conditions. Moreover, we 
also include the values of the deceleration parameter, calculated through 
(\ref{deccelmodIIsol2b}). Each of the above critical points have contracting partners via 
the discrete symmetry \eqref{discrete_symm_II}, which are displayed in Table \ref{Tab4b}. 

Lastly, there exists a line of static solutions, i.e. neither expanding nor contracting, 
given by 
\begin{equation}
S_1: \left(T,Q_0,\Theta_{DE}\right)=\left(T_c,0, \frac{(T_c-1) (3 w_m+1)}{3 T_c (w_m-1)-3 
(w_m+1)}\right), \quad T_c\in [0,1].
\end{equation}
Imposing the physical condition $0\leq w_m\leq 1$, it follows  that the above line always 
satisfies the existence condition $0\leq \Theta_{DE}\leq 1$.
The eigenvalues of the linearization around $S_1$ are
\begin{align*}
0,
-\frac{\sqrt{(3 w_{m}+1) \left(1-T_c\right) \left[(w_{m}-1) T_c-w_{m}-1\right] \left[(2 
w_{m}+1) T_c^2-(w_{m}+6)
		T_c-w_{m}-1\right]}}{(w_{m}-1) T_c-w_{m}-1},\\
\frac{\sqrt{(3 w_{m}+1) \left(1-T_c\right) \left[(w_{m}-1) T_c-w_{m}-1\right]
		\left[(2 w_{m}+1) T_c^2-(w_{m}+6) T_c-w_{m}-1\right]}}{(w_{m}-1) 
T_c-w_{m}-1},
\end{align*} 
and thus whenever this line exists these eigenvalues are always real. Since 
two of them have different sign the whole line behaves as saddle.

We mention that  in order to determine the stability of the 
six nonhyperbolic critical points (expanding and contracting ones) we apply the center 
manifold 
method \cite{wiggins}, 
and the corresponding analysis is performed in Appendix \ref{App3}. 
 
\begin{table*}
  \resizebox{\columnwidth}{!}
{
\!\!\!\!\!\!\!\!\!\!\!
	\begin{tabular}{cccccccc}
		\hline  
		Label & $Q_0$& $\Theta_{DE}$ & $T$ & $q$ & Existence 
& 
Eigenvalues &  \!\!\!\!\!\!\!\! Stability \\ 
		\hline  
		$Q_{13}$ & $1$ & $0$ & $0$ &$\frac{3w_m+1}{2}$ & always & {\small{ $3, 
3(1+w_m), 1+3 w_m$}} &  \!\!\!\!\!\!\!\! unstable \\ 
		\hline
		$Q_{14}$ & $1$ & $1$ & $0$ &$-1$& always & {\small{$3, 
-3(1+w_m), -2$}} & \!\!\!\!\!\!\!\!  saddle \\ 
		\hline
		$Q_{15}$ & $1$ & $0$ & $1$ &{\small{arbitrary}}& always & $6, 0, 0$ & 
\!\!\!\!\!\!\!\! {\small{nonhyperbolic,}}
  			\\
			&&&&&&&  \!\!\!\!\!\!\!\! {\small{behaves as saddle}} \\
		\hline
		$Q_{16}$ & $Q_{0 c}$ & $0$ & $1$ & {\small{arbitrary}} & 
$Q_{0 c}\in (0,1]$ & $6 Q_{0 c}, 0, 
0$ & \!\!\!\!\!\!\!\!\!\! {\small{ nonhyperbolic,}}
  			\\
			&&&&&&&  \!\!\!\!\!\!\!\!  {\small{behaves saddle}} \\
		\hline
		$Q_{17}$ & $1$ & $1$ & $1$&$-\infty$ & always & $-6, -6, 0$ & 
\!\!\!\!\!\!\!\! {\small{nonhyperbolic,}}
  			\\
			&&&&&&&  \!\!\!\!\!\!\!\!   {\small{behaves as 
stable }}\\
		\hline
	\end{tabular} }
	\caption{\label{Tab4}
	 The physical critical points of the system   \eqref{systII4} of 
time asymmetric cosmology of Model II corresponding to expanding cosmologies:  $f(V)=g_2 
e^{\lambda 
V}$, with positive 
curvature and $\lambda>0$, and their existence and stability conditions. We assume $0\leq 
w_m\leq 
1$.}
\end{table*}

 \begin{table*}
 	\resizebox{\columnwidth}{!}
 	{
 		\!\!\!\!\!\!\!\!\!\!\!
 		\begin{tabular}{cccccccc}
 			\hline  
 			Label & $Q_0$& $\Theta_{DE}$ & $T$ & $q$ & Existence 
 			& 
 			Eigenvalues &  \!\!\!\!\!\!\!\! Stability \\ 
 			\hline  
 			$R_{13}$ & $-1$ & $0$ & $0$ &$\frac{3w_m+1}{2}$ & always & 
{\small{ $-3, 
 					-3(1+w_m), -(1+3 w_m)$}} &  \!\!\!\!\!\!\!\! 
stable \\ 
 			\hline
 			$R_{14}$ & $-1$ & $1$ & $0$ &$-1$& always & {\small{$-3, 
 					3(1+w_m), 2$}} & \!\!\!\!\!\!\!\!  saddle \\ 
 			\hline
 			$R_{15}$ & $-1$ & $0$ & $1$ &{\small{arbitrary}}& always & $-6, 
0, 
0$ & 
 			\!\!\!\!\!\!\!\! {\small{nonhyperbolic,}}
 			\\
 			&&&&&&&  \!\!\!\!\!\!\!\! {\small{behaves as 
 					saddle}} \\
 			\hline
 			$R_{16}$ & $-Q_{0 c}$ & $0$ & $1$ & {\small{arbitrary}} & 
 			$Q_{0 c}\in (0,1]$ & $-6 Q_{0 c}, 0, 
 			0$ & \!\!\!\!\!\!\!\!\!\! {\small{ nonhyperbolic,}}
 			\\
 			&&&&&&&  \!\!\!\!\!\!\!\!  {\small{behaves as saddle}} \\
 			\hline
 			$R_{17}$ & $-1$ & $1$ & $1$&$-\infty$ & always & $6, 6, 0$ & 
 			\!\!\!\!\!\!\!\! {\small{nonhyperbolic,}}
 			\\
 			&&&&&&&  \!\!\!\!\!\!\!\!   {\small{behaves as 
 					unstable }}\\
 			\hline
 		\end{tabular} }
 		\caption{\label{Tab4b}
 			The physical critical points of the system   \eqref{systII4} of 
 			time asymmetric cosmology of Model II corresponding to 
contracting 
cosmologies:  $f(V)=g_2 e^{\lambda V}$, with positive 
 			curvature and $\lambda>0$, and their existence and stability 
conditions. We assume $0\leq w_m\leq 1$.}
 	\end{table*}

\item  $\lambda<0$

In this case we define the additional auxiliary variable
 \begin{align}
\label{auxmodIITbb}
{T_{1}}=-\frac{\lambda a^3}{1-\lambda a^3}.
\end{align}
Since $0< {T_{1}}< 1$ we define  
\begin{equation}
\frac{d\check{\tau}}{dt}=\mathcal{D}(1-{T_{1}})^{-1},
\end{equation}
so that $\check{\tau}$ is monotonic increasing for  $\mathcal{D}(1-{T_{1}})^{-1}>0$. 
Therefore, 
using the 
auxiliary variables (\ref{varsIIa}) and (\ref{auxmodIITbb}) we can re-write the 
cosmological equations in autonomous form
as
\begin{subequations}
	\label{systII4b}
	\begin{align}
     &\frac{d {T_{1}}}{d\check{\tau}}=3 Q_0 {T_{1}}(1-{T_{1}})^2,\\
     &\frac{d Q_0}{d\check{\tau}}=-\frac{1}{2} \left(1-Q_0^2\right) \left[3 ({T_{1}}-1) 
(w_m+1) (\Theta_{DE}-1)+2 (3 {T_{1}} \Theta_{DE}+{T_{1}}-1)\right],\\
     &\frac{d \Theta_{DE}}{d\check{\tau}}=3 Q_0 (\Theta_{DE}-1) 
\Theta_{DE} ({T_{1}} (w_m+3)-w_m-1),
	\end{align}
\end{subequations}
where we have used the constraint (\ref{constrII2}) in order to eliminate $\Omega_m$. 
The above system is defined on the 
$\left\{({T_{1}},Q_0, \Theta_{DE}): 0\leq {T_{1}} \leq 1,  -1\leq 
Q_0\leq 1, 0\leq \Theta_{DE}\leq 1\right\}$ part of the phase space, 
where we 
have attached the invariant boundaries ${T_{1}}=0$ and ${T_{1}}=1$. 

Similarly to the previous section, the function   
\begin{equation}
\check{N}=\frac{{T_{1}}}{1-{T_{1}}}, \quad \check{N}'=3 (1-{T_{1}}) Q_0 \check{N},
\end{equation}  
is a monotonic function in the region $Q_0<0, 0<{T_{1}}<1$, where $\check{N}$ is 
monotonically increasing, and in the region $Q_0>0, 0<{T_{1}}<1$, where $\check{N}$ is 
monotonically decreasing, and it follows 
that either ${T_{1}}\rightarrow 1$ or ${T_{1}}\rightarrow 0$ asymptotically. 

Furthermore, the system \eqref{systII4b}   is invariant under the symmetry 
\begin{equation}
\label{discrete_symm_IIb}
\left(\check{\tau}, {T_{1}}, Q_0,  \Theta_{DE}\right)\rightarrow \left(-\check{\tau}, 
{T_{1}}, -Q_0,
  \Theta_{DE}\right).
\end{equation}
Thus, it is sufficient to discuss the behavior in one part of the phase space, that is in
$\check{\tau}
\geq 0, {T_{1}}\geq 0, Q_0\geq 0, \Theta_{DE} \geq 0$, while the dynamics on the other 
part is being obtained from \eqref{discrete_symm_IIb}. For example, if a point with 
coordinates 
$({T_{1}}^*, Q_0^*
,\Theta_{DE}^*), Q_0^*>0, \Theta_{DE}^*>0$ is a future attractor as 
$\check{\tau}\rightarrow +\infty$, then, its partner point 
$({T_{1}}^*,-Q_0^*,\Theta_{DE}^*)$ via \eqref{discrete_symm_IIb} is 
a past attractor as $\check{\tau}\rightarrow -\infty$, and vice versa. Additionally, in 
the invariant set ${T_{1}}=0$ we obtain the first integral 
\begin{equation}
\frac{(1-Q_0^2)^{3(1+w_m)}}{(1- \Theta_{DE})^2 \Theta_{DE}^{1+3 w_m}}= c,
\end{equation}
with $c$ an integration constant. 

Finally, note that 
in terms of the auxiliary variables (\ref{varsIIa}),(\ref{auxmodIITbb}) the deceleration 
parameter (\ref{deccelmodII}) is given as 
\begin{equation}
 q=  \frac{1}{2Q_0^2}\left[1+3w_m 
(1-\Theta_{DE}) 
-3\left(\frac{1-3{T_{1}}}{1-{T_{1}}}\right)\Theta_{DE}\right].
\label{deccelmodIIsol2bbb}
\end{equation}
Hence, for the critical points having ${T_{1}}=0$, the expression 
\eqref{deccelmodIIsol2bbb} is 
well-defined 
and leads to $
q\left|_{{T_{1}}=0}\right. = \frac{1}{2Q_0^2}\left[1+3w_m 
(1-\Theta_{DE}) -3 \Theta_{DE}\right]$, however  for the critical points having 
${T_{1}}=1, \Theta_{DE}\neq 0$, it follows that 
$q\rightarrow +\infty$ as  ${T_{1}}\rightarrow 1^-$.
On the other hand, for the critical points having ${T_{1}}=1,\Theta_{DE}=0$, 
$q$ is arbitrary.

$\ \ \,$The scenario at hand admits four physical critical points, and one curve of 
critical points (namely $Q_{21}$) representing accelerating solutions ($Q_0>0$), which 
are displayed in Table \ref{Tab4IIa} along with their existence conditions. In the same 
Table we present the eigenvalues of the corresponding perturbation matrix, and the 
resulting stability conditions. Finally, we also include the values of the deceleration 
parameter, calculated through (\ref{deccelmodIIsol2bbb}). Each point/curve in Table 
\ref{Tab4IIa} has a partner through the symmetry \eqref{discrete_symm_IIb}, representing 
a contracting cosmology ($Q_0<0$), which are displayed in Table \ref{Tab4IIb}.

Lastly, the system \eqref{systII4b} admits a line representing static solutions, i.e. 
neither expanding nor contracting,  given by 
\begin{equation}
S_2: \left({T_{1}}, Q_0,  \Theta_{DE}\right)=\left(T_c, 0, \frac{(3 w_m+1) 
\left(T_c-1\right)}{3 \left(w_m T_c+3 T_c-w_m-1\right)}\right),
\end{equation}
with eigenvalues
\begin{align*}
&0, -\frac{\sqrt{T_c-1} \sqrt{(3 w_m+1) \left[(w_m+3) T_c-w_m-1\right] \left[(4 w_m+15) 
T_c^2-5 (w_m+2)
		T_c+w_m+1\right]}}{(w_m+3) T_c-w_m-1}, \\
&\frac{\sqrt{T_c-1} \sqrt{(3 w_m+1) \left[(w_m+3) T_c-w_m-1\right] \left[(4
		w_m+15) T_c^2-5 (w_m+2) T_c+w_m+1\right]}}{(w_m+3) T_c-w_m-1}.
\end{align*}
This line exists and is physical, i.e. possessing $0\leq \Theta_{DE} \leq 1$,  for 
i) $0\leq T_c<\frac{5 w_m-\sqrt{3 w_m (3 w_m+8)+40}+10}{8 w_m+30}$, when the eigenvalues 
are real and $S_2$ behaves as saddle, or
ii) $\frac{5 w_m-\sqrt{3 w_m (3 w_m+8)+40}+10}{8 w_m+30}<T_c\leq \frac{1}{4}$, when two 
eigenvalues are purely imaginary.

 We mention that  in order to determine the stability of the 
six nonhyperbolic critical points we apply the center manifold method \cite{wiggins}, 
and the corresponding analysis is performed in Appendix \ref{App4}.

\begin{table*}
  \resizebox{\columnwidth}{!}
{
\!\!\!\!\!\!\!\!\!\!\!
	\begin{tabular}{cccccccc}
		\hline  
		Label & $Q_0$& $\Theta_{DE}$ & ${T_{1}}$ & $q$&
Existence & 
		Eigenvalues & \!\!\!\!\!\!Stability \\ 
		\hline  
		$Q_{18}$ & $1$ & $0$ & $0$ &$\frac{3w_m+1}{2}$ & always & {\small{$3, 
3(1+w_m), 1+3 w_m$ }}& \!\!\!\!\!\!unstable \\ 
		\hline
		$Q_{19}$ & $1$ & $1$ & $0$& $-1$ & always & {\small{$3, 
-3(1+w_m), -2$}} & \!\!\!\!\!\!saddle \\ 
		\hline
		$Q_{20}$ & $1$ & $0$ & $1$ &  {\small{arbitrary}} & always & $-6, 0, 0$ 
& 
\!\!\!\!\!\!{\small{nonhyperbolic,}}
  			\\
			&&&&&&&\!\!\!\!\!\! {\small{behaves as saddle}} \\ 
		\hline
		$Q_{21}$ & $Q_{0 c}$ & $0$ & $1$ & {\small{arbitrary}} & 
$Q_{0 c}\in 
(0,1)$ & $-6 Q_{0 c}, 0, 
		0$ &\!\!\!\!\!\!{\small{nonhyperbolic,}}
  			\\
			&&&&&&&\!\!\!\!\!\! {\small{behaves as saddle}}   \\ 
		\hline
		$Q_{22}$ & $1$ & $1$ & $1$ & $+\infty$ & always & $6, 6, 0$ & 
\!\!\!\!\!\!{\small{nonhyperbolic,}}
  			\\
			&&&&&&&  \!\!\!\!\!\!{\small{ behaves as saddle}}\\ 
		\hline
	\end{tabular} }
	\caption{\label{Tab4IIa} 
	 The physical critical points of the system   \eqref{systII4b} of 
time asymmetric cosmology of Model II representing expanding cosmologies:  $f(V)=g_2 
e^{\lambda V}$,
 with positive 
curvature and $\lambda<0$, and their existence and stability conditions. We assume $0\leq 
w_m\leq 
1$.}
\end{table*}

\begin{table*}
	\resizebox{\columnwidth}{!}
	{
		\!\!\!\!\!\!\!\!\!\!\!
		\begin{tabular}{cccccccc}
			\hline  
			Label & $Q_0$& $\Theta_{DE}$ & ${T_{1}}$ & $q$&
			Existence & 
			Eigenvalues & \!\!\!\!\!\!Stability \\ 
			\hline  
			$R_{18}$ & $-1$ & $0$ & $0$ &$\frac{3w_m+1}{2}$ & always & 
{\small{$-3, 
					-3(1+w_m), -(1+3 w_m)$ }}& \!\!\!\!\!\!stable \\ 
			\hline
			$R_{19}$ & $-1$ & $1$ & $0$& $-1$ & always & {\small{$-3, 
					3(1+w_m), 2$}} & \!\!\!\!\!\! saddle \\ 
			\hline
			$R_{20}$ & $-1$ & $0$ & $1$ &  {\small{arbitrary}} & always & $6, 
0, 0$ 
			& 
			\!\!\!\!\!\!{\small{nonhyperbolic,}}
			\\
			&&&&&&&\!\!\!\!\!\! {\small{ behaves as saddle}} \\ 
			\hline
			$R_{21}$ & $-Q_{0 c}$ & $0$ & $1$ & {\small{arbitrary}} & 
			$Q_{0 c}\in 
			(0,1)$ & $6 Q_{0 c}, 0, 
			0$ &\!\!\!\!\!\!{\small{nonhyperbolic,}}
			\\
			&&&&&&&\!\!\!\!\!\! {\small{behaves as saddle}}   \\ 
			\hline
			$R_{22}$ & $-1$ & $1$ & $1$ & $+\infty$ & always & $-6, -6, 0$ & 
			\!\!\!\!\!\!{\small{nonhyperbolic,}}
			\\
			&&&&&&&  \!\!\!\!\!\!{\small{behaves as saddle}}\\ 
			\hline
		\end{tabular} }
		\caption{\label{Tab4IIb} 
			The physical critical points of the system   \eqref{systII4b} of 
			time asymmetric cosmology of Model II representing contracting 
cosmologies:  $f(V)=g_2 e^{\lambda V}$, with positive 
			curvature and $\lambda<0$, and their existence and stability 
conditions. We assume $0\leq w_m\leq 1$.}
	\end{table*}

\end{itemize}

In summary, the scenario  of Model II, namely  $f(V)=g_2 e^{\lambda V}$, with zero or 
negative curvature admits the following stable late-time solutions:
For $\lambda>0$ the contracting solution $R_{13}$  and 
the expanding solution $Q_{17}$ (nonhyperbolic but with stable center manifold), while 
for $\lambda<0$ the contracting solution $R_{18}$.

  \section{Physical Implications} 
\label{PhysImplic}   
   
Having performed a complete dynamical analysis of cosmological scenarios governed by time 
asymmetric extensions of general relativity , we can now proceed to the discussion of 
the physical implications. In particular, we focus on the stable late-time solutions, 
since these solutions can attract the universe at late times, independently of the 
specific initial conditions and the specific intermediate evolution.

 \subsection{Model I:  $f(V)=g_1V^{m}$}
\label{ModIanalysisphysimpl}

In the case where $f(V)=g_1V^{m}$, i.e. when $g(a)=\frac{g_1}{G}a^p$  (with $p=3m+1$), 
with $g_1$ a constant and $p$ a parameter, and with open or zero curvature, the scenario 
at hand exhibits the three critical points presented in Table \ref{Tab1}. Point $P_1$ 
corresponds to a dark-matter dominated universe ($\Omega_m=1$), that is non-accelerating 
($q>0$), however it is never stable and thus it cannot attract the universe at late 
times. Point $P_2$ corresponds to a universe governed by the curvature term 
($\Omega_k=1$), which is neither accelerating nor decelerating (this is typical for 
curvature dominated solutions \cite{Copeland:2009be}). For $p<0$ it can be stable, and 
thus it can 
attract the universe at late times (this is actually expected since for $p<0$ the 
effective dark-energy term decreases faster than the curvature term, and hence the latter 
dominates). However, its observational features are disfavored by observations. Point 
$P_3$ is stable for $p>0$ and thus it can be the stable late-time state of the 
universe. It corresponds to a dark-energy dominated, accelerating universe, where the 
dark-energy equation-of-state parameter (\ref{wdeI}), namely $w_{DE} =-(1+2p)/3$, can lie 
either in the quintessence regime (for $0<p<1$), or in the phantom one (for $1<p$), or  
either behave as an effective cosmological constant (for $p=1$) giving rise to a de 
Sitter universe. These features make it a good candidate for the description of the 
universe, especially if $0.9\lesssim p\lesssim 1.1$, in which case $-1.07\lesssim 
w_{DE}\lesssim -0.93 $ in agreement with observations \cite{Planck:2015xua}. We mention 
that the above behavior is obtained without the addition of an explicit cosmological 
constant term in the action, i.e. it is a pure effect of the novel, time-asymmetric 
theory. Finally, note that even when the effective dark energy lies in the phantom 
regime, the universe does not end in a Big Rip 
\cite{Caldwell:2003vq,Sami:2003xv,Hao:2004ky,Nojiri:2015fia}, or any other type of 
singularity \cite{Nojiri:2005sx}, at finite time.
\begin{figure}[ht]
\centering
\includegraphics[width=0.6\linewidth]{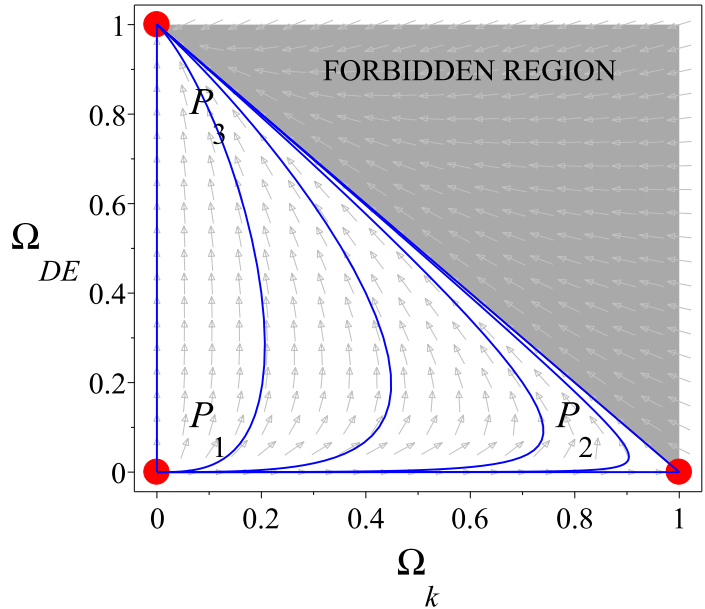}
\caption{{\it{The phase-space behavior of time asymmetric cosmology 
of Model I:  $f(V)=g_1V^{m}$, with negative curvature, $p=0.9$ (i.e. $m=-0.033$), and 
$w_m=0$. 
The shadowed region marks the  unphysical part of the phase space. In
this specific example the universe is led to the the  dark-energy dominated, accelerating
solution $P_3$.}}}
\label{fig:Fig8}
\end{figure}

In order to present the above behavior in a more transparent way, we evolve numerically  
the cosmological equations and in Fig. \ref{fig:Fig8} we depict the corresponding 
phase-space behavior. The unphysical part of the phase space (in which the density 
parameters exceed one) is marked by the shadowed region. As we can see, in this 
specific example the universe results in the  dark-energy dominated, accelerating
solution $P_3$.
 \begin{figure}[ht]
 	\centering
 	\includegraphics[width=0.6\linewidth]{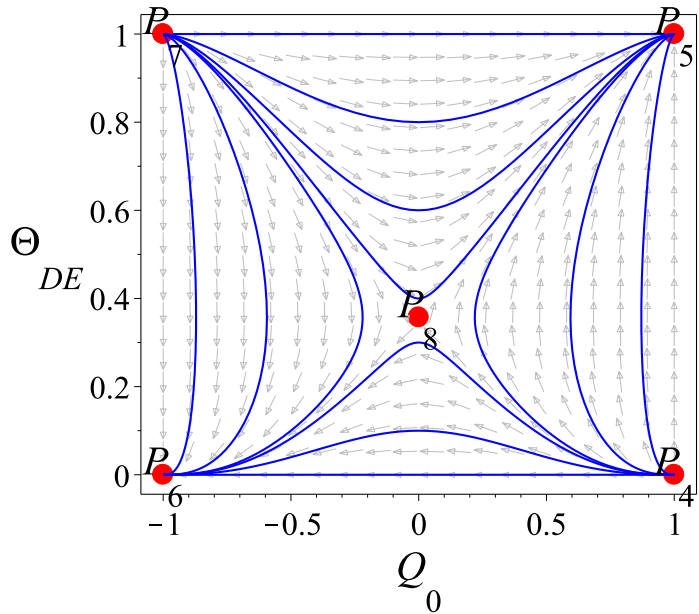}
 	\caption{{\it{The phase-space behavior of time asymmetric cosmology 
 				of Model I:  $f(V)=g_1V^{m}$, with positive curvature, 
$p=0.9$ (i.e. $m=-0.033$), and 
 				$w_m=0$. In this specific example the universe is led to 
either
(a) the dark-energy dominated,  accelerating
 				solution $P_5$ or (b) the matter dominated,  
contracting solution $P_6$.}}}
 	\label{fig:Fig8b}
 \end{figure}

In the case of positive curvature, the model possesses five critical points, displayed in 
Table \ref{Tab2}. Amongst them, the points $P_5$ and $ P_6$ can be stable, and thus they 
can attract the universe at late times. $P_5$ corresponds to an accelerating, 
dark-energy dominated universe ($\Omega_{DE}=\Theta_{DE}=1$ since for this point $ 
\mathcal{D}\rightarrow H$ in (\ref{auxmodIB})), in which the dark-energy 
equation-of-state parameter can lie either in the quintessence or in the phantom regime, 
or behave like an effective cosmological constant. Hence, it can be a good candidate for 
the description of the universe. On the other hand, $P_6$ corresponds to a matter 
dominated, contracting solution, and as we mentioned before it could be an attractor 
too, but it does not describe accurately the universe at late times. 
Moreover, point $P_7$ has the reverse dynamical behavior of $P_5$  due to 
\eqref{discrete_symm}, i.e it corresponds to a contracting ($Q_0<0$), dark-energy 
dominated universe ($\Omega_{DE}=\Theta_{DE}=1$), and it can be an attractor too. 
Similarly, point $P_4$   presents the time reversal behavior of $P_6$. Note that 
there exist orbits connecting $P_4$ and $P_5$  
with  $P_6$ and $ P_7$, which implies that $Q_0$ can cross zero, i.e. $H=0$, during the 
evolution, and thus the universe exhibits a bounce or a cosmological turnaround. Finally, 
the system admits an static solution, $P_8$, which always behaves as a saddle point, hence
it cannot represent the late-time universe. We mention that the above features are not 
obtained in the flat or open curvature, where $H$ cannot change sign. 

 In order to present the above behavior in a more transparent way, we evolve numerically  
 the cosmological equations and in Fig. \ref{fig:Fig8b} we depict the corresponding 
 phase-space behavior. As we can see, in this 
 specific example the universe results in the  dark-energy dominated, accelerating
 solution $P_5$ or in the matter dominated, contracting solution $P_6$.

\subsection{Model II:   $f(V)=g_2 e^{\lambda V}$}
\label{ModIIanalysisphysimpl}

\begin{figure}[ht]
\centering
\includegraphics[width=0.6\linewidth]{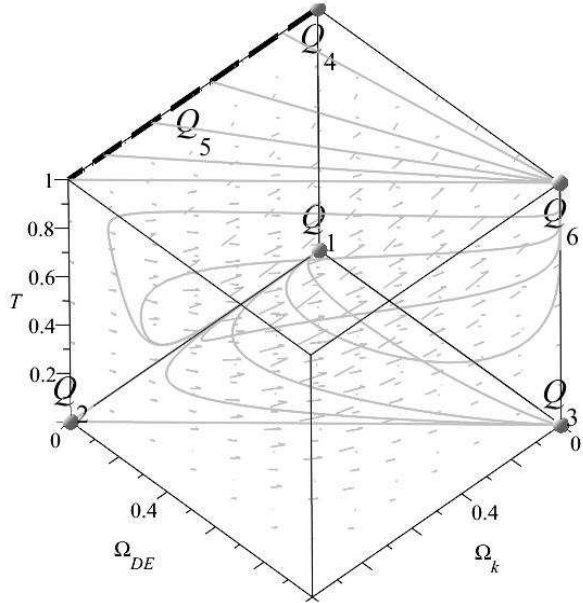}
\caption{{\it{ The phase-space behavior of time asymmetric cosmology 
of Model II:  $f(V)=g_2 e^{\lambda V}$, with negative curvature, 
$w_m=0$ and $\lambda>0$ (the specific value of $\lambda$ is not relevant, only its sign, 
since it has 
been absorbed into the auxiliary variable $T$ according to (\ref{auxmodIIT})). In this 
specific example the universe is led to the  dark-energy dominated, 
accelerating solution $Q_6$. The bold dashed line named $Q_5$ in general presents saddle 
behavior, however it is a local source for all the orbits (which have the shape of 
straight lines connecting it with $Q_6$) located at the invariant set $T=1$.}}}
\label{figQ6}
\end{figure}

In the case where $f(V)=g_2 e^{\lambda V}$, i.e. when $g(a)=g_2\frac{a}{G} 
e^{\lambda a^3}$, with $g_2$ a constant and $\lambda$ a parameter, with open or zero 
curvature, and $\lambda>0$, the scenario at hand exhibits five isolated critical points 
and one curve of critical points presented in Table \ref{Tab3}. Amongst them, only point 
$Q_6$ behaves like a stable one (although nonhyperbolic) and thus it can be the late-time 
state of the universe. It corresponds to a dark-energy  dominated universe, in which the 
dark-energy equation-of-state parameter lies in the phantom regime. Note however that as 
the universe approaches this point, the deceleration parameter $q$ decreases 
monotonically, resulting to a divergence at the critical point. In particular, as the 
scale factor increases and the dark energy term becomes dominant, we can obtain an 
approximate solution for the scale factor, namely the inverse of  
$t-t_0=\frac{\text{Ei}\left(-a^3 \lambda \right)}{3 g_2}=e^{-a^3 \lambda } {\mathcal{
O}}\left(\left(\frac{1}{a}\right)^3\right)$, where $t_0=-c_1/g_2$, with 
$\text{Ei}(z),z<0$, the exponential integral function and  $c_1$ an 
integration 
constant, and we can immediately see that the scale factor diverges at a finite time, 
which is the realization of a Big Rip \cite{Nojiri:2005sx}. This behavior was 
expected, 
since for 
$\lambda>0$ the extra, time-asymmetric, term that constitutes the effective dark energy 
sector increases monotonically. Hence, for these parameter choices, the scenario at hand 
does not correspond to the usual classes of cosmological models, and thus it should not 
be considered as a successful one. In Fig. \ref{figQ6} we depict the phase-space 
behavior of such a scenario, arising from numerical elaboration. As we observe, in this 
example the universe results in the  dark-energy dominated, accelerating
solution $Q_6$.

\begin{figure}[ht]
\centering
\includegraphics[width=0.7\linewidth]{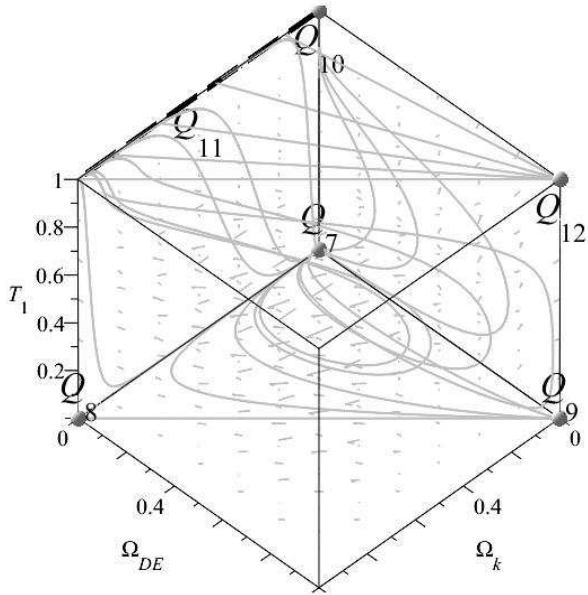}
\caption{{\it{ The phase-space behavior of time asymmetric cosmology 
of Model II:  $f(V)=g_2 e^{\lambda V}$, with negative curvature,  
$w_m=0$ and $\lambda<0$ (the specific value of $\lambda$ is not relevant, only its sign, 
since it has been absorbed into the auxiliary variable ${T_{1}}$ according to 
(\ref{auxmodIITb})). In this specific example the universe is led to the dark-matter 
dominated solution $Q_{10}$ (if $\Omega_k=0$ at the initial state), or to the 
curvature-dominated solution located on one endpoint of line $Q_{11}$, namely 
$(\Omega_k,\Omega_{DE}, {T_{1}})=(1,0,1)$ (if $\Omega_k>0$ at the initial state). All 
other points of the curve $Q_{11}$, which is represented by a bold dashed line, behave as
saddle.}}}
\label{Fig3}
\end{figure}

In the case of zero or open curvature and $\lambda<0$, the model exhibits five 
isolated critical points and one curve of critical points, displayed in Table 
\ref{Tab3b}. Amongst them, point $Q_{10}$ behaves as stable for the flat models, and thus 
it can attract the universe at late times. However, it corresponds to a  dark-matter 
dominated universe, and thus it is not favored by observations. This was expected, since 
for $\lambda<0$ the effective dark-energy terms are redshifted away in a much faster way 
(due to the exponential) than the matter contribution, leaving the universe matter 
dominated. Nevertheless, one could improve this behavior by the addition of an explicit 
cosmological constant, in which case he could get the correct thermal history, namely the 
succession of matter and dark-energy eras. However, since in this work we are interested 
in investigating the effects of the pure time-asymmetric cosmology, without the explicit 
presence of a cosmological constant, we do not  examine such a possibility further. 
Additionally, as we describe in detail in Appendix \ref{App2}, the nonhyperbolic curve of 
critical points $Q_{11}$, with the exception of its endpoint with $\Omega_k=1$,  behaves 
as saddle. For $0<\Omega_k<1$, it corresponds to a universe with $\Omega_{DE}=0$, however 
not completely matter-dominated, since the curvature contribution remains non-zero. 
Another interesting point located on the curve $Q_{11}$ is the one corresponding to 
complete curvature domination, namely with $\Omega_k=1$. This point is indeed a stable 
late-time state of the universe. Similarly to $Q_{10}$, the above features are not favored 
by observations to be the late-time state of the universe, however these curves of points 
could be a good candidate for the description of its intermediate phases, especially under 
the addition of an explicit cosmological constant. In Fig. \ref{Fig3}, through a 
numerical elaboration, we present the phase-space behavior of this model. As we see, in 
this example if the universe starts with  $\Omega_k=0$ it results in the  dark-matter 
dominated solution $Q_{10}$. On the other hand, if $\Omega_k>0$ initially then the 
universe results in the curvature-dominated solution $(\Omega_k,\Omega_{DE}, 
{T_{1}})=(1,0,1)$ located on the bold dashed line $Q_{11}$.

\begin{figure}[ht]
\centering
\includegraphics[angle=90,width=0.8\linewidth]{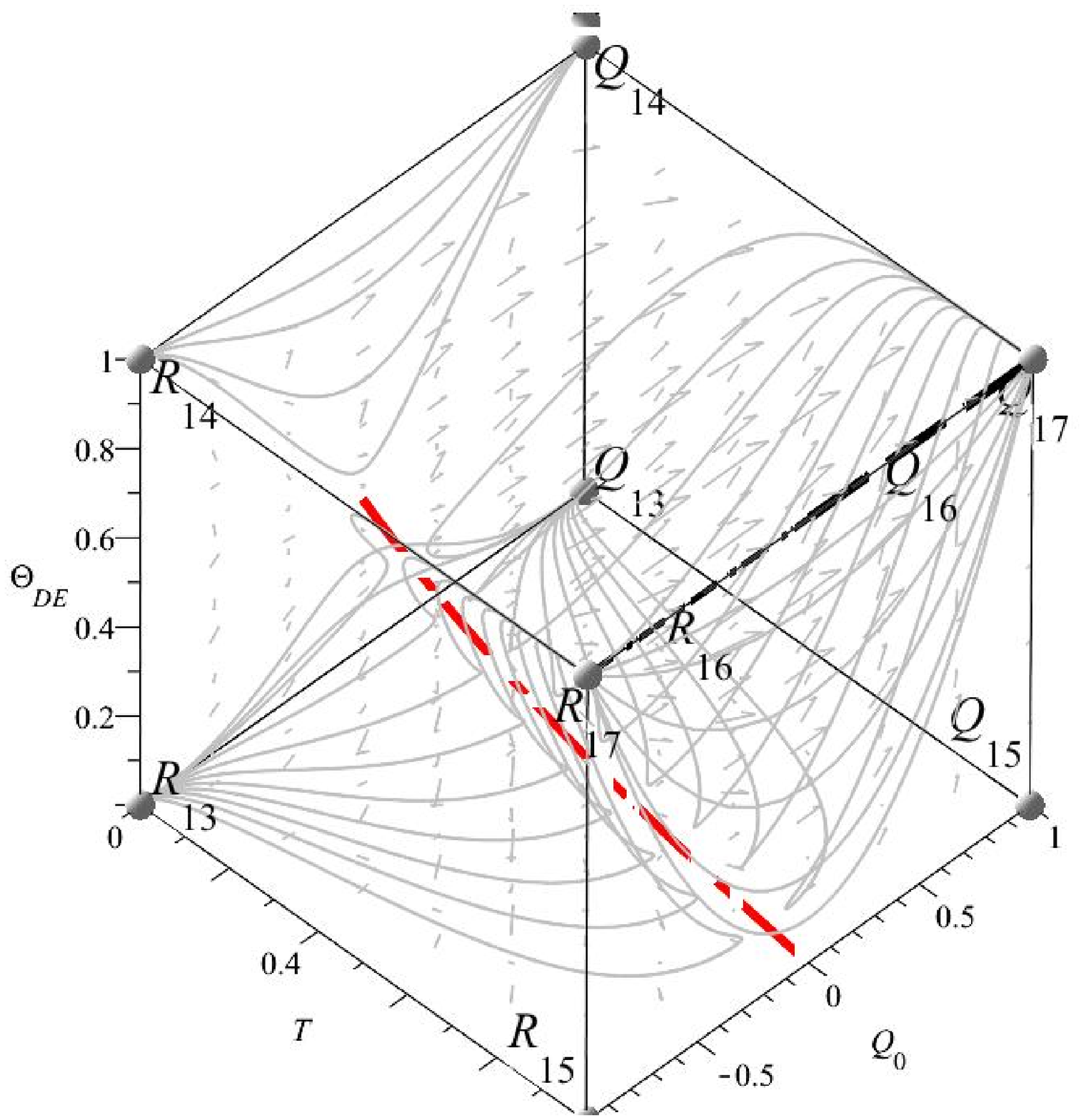}
\caption{{\it{ The phase-space behavior of time asymmetric cosmology 
of Model II:  $f(V)=g_2 e^{\lambda V}$, with positive curvature,    
$w_m=0$ and $\lambda>0$ (the specific value of $\lambda$ is not relevant, only its sign, 
since it has been absorbed into the auxiliary variable $T$ according to 
(\ref{auxmodIIT})). In this specific example the universe is led to either the dark-energy 
dominated, accelerating solution $Q_{17}$, or to the contracting solution $R_{13}$. The 
dot-dashed (red) line represents the curve of static solutions $S_1$. Notice the 
presence of orbits crossing the line $Q=0$, i.e. $H=0$, which correspond to
transitions from expanding to contracting cosmologies and vice versa, that is to 
cosmological turnarounds and bounces.}}}
\label{Fig4}
\end{figure}

In the case of positive curvature and $\lambda>0$ the scenario at hand exhibits four 
physical critical points, and one curve of critical points, namely $Q_{16}$, which is the 
straight line joining the points $(0,0,1)$ and $(1,0,1)$ (with the left endpoint not 
included), corresponding to expanding cosmologies, which are summarized in Table 
\ref{Tab4}. All these critical points have contracting partners via the 
discrete symmetry \eqref{discrete_symm_II}, which are displayed in Table \ref{Tab4b}. 
Additionally, there exists a line of static solutions namely $S_1$, however since they are 
saddle they cannot attract the universe at late times. Amongst all these points, the 
late-time attractors are the expanding solution $Q_{17}$ and the contracting $R_{13}$. 
In particular, $Q_{17}$ corresponds to a dark-energy dominated universe in which the 
dark-energy equation-of-state parameter is phantom-like. Note however that as the 
universe approaches this point, the deceleration parameter $q$ decreases monotonically, 
resulting to a divergence at the critical point. Using similar arguments as for point 
$Q_6$ for the open or zero curvature case, it can be shown that it is of a finite-time 
type, namely a Big Rip \cite{Nojiri:2005sx}. Similarly to the open or zero curvature case, 
this behavior was expected, since for $\lambda>0$ the extra, time-asymmetric, term that 
constitutes the effective dark energy sector increases monotonically. Additionally, there 
is another stable late-time solution, namely the matter-dominated point $R_{13}$ which 
ends in a Big-Cruch. In Fig. \ref{Fig4} we depict the phase-space behavior of this 
scenario. As we see, in this example the universe results in the  dark-energy dominated, 
accelerating solution $Q_{17}$ or in the Big-Crunch singularity $R_{13}$. 

\begin{figure}[ht]
\centering
\includegraphics[angle=90,width=0.6\linewidth]{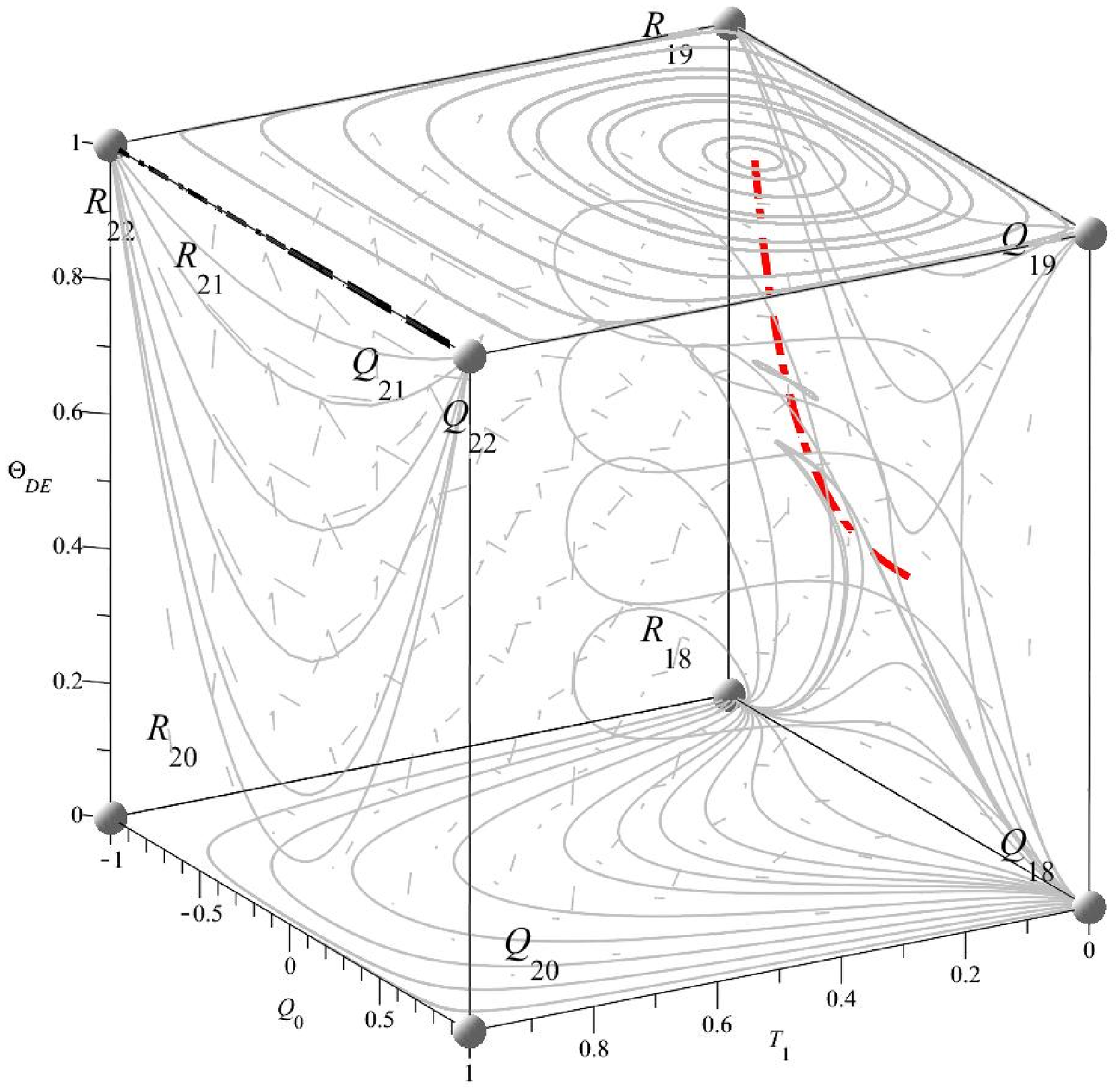}
\caption{{\it{ The phase-space behavior of time asymmetric cosmology 
of Model II:  $f(V)=g_2 e^{\lambda V}$, with positive curvature,    
$w_m=0$ and $\lambda<0$ (the specific value of $\lambda$ is not relevant, only its sign, 
since it has been absorbed into the auxiliary variable ${T_{1}}$ according to 
(\ref{auxmodIITb})). In this specific example the universe is led to the contracting 
solution $R_{18}$. Additionally, the figure shows orbits exhibiting the crossing of the 
$Q_0=0$ line, which correspond the transition from contracting to expanding cosmologies 
and vice versa.
}}}
\label{Fig5}
\end{figure}

In the case of positive curvature and $\lambda<0$, the model exhibits four isolated 
critical points and one curve of critical points,  corresponding to expanding 
cosmologies, displayed in Table \ref{Tab4IIa}. Each of the above critical points have 
contracting partners via the discrete symmetry \eqref{discrete_symm_II}, which are 
displayed in Table \ref{Tab4IIb}. Amongst them, point $R_{18}$ behaves as stable, and thus 
it can be the late-time state of the universe. However, it corresponds to a contracting  
dark-matter dominated universe, and therefore it is not favored by observations. Similarly 
to the open or flat case, this was expected since for $\lambda<0$ the effective 
dark-energy terms are redshifted away in a much faster way  than the matter contribution. 
Additionally, there exists a line of static solutions which are saddle, while, as we 
describe in detail in Appendix \ref{App4}, the nonhyperbolic curves of 
critical points $Q_{20}$ and $Q_{21}$ behave typically as saddle. 
In Fig. \ref{Fig5} we present the phase-space behavior for the model at hand, where we 
observe that the late-time attractor is the contracting solution $R_{18}$. Additionally, 
the figure shows orbits exhibiting the crossing of the $Q_0=0$ line, which correspond the 
transition from contracting to expanding cosmologies and vice versa.

  \section{Conclusions} 
\label{Conclusions}

In this work we studied the cosmological behavior in a universe governed by time 
asymmetric extensions of general relativity. This novel modified gravity is based on the 
addition on the Hamiltonian framework of new, time-asymmetric, terms, in a way that the 
algebra of constraints and local physics remain unchanged \cite{Cortes:2015ola}. However, 
at cosmological scales these new terms can have significant effects that can alter the 
universe evolution, both at early and late times. In particular, assuming that the new 
terms in the Hamiltonian are proportional to an arbitrary function of the spatial volume, 
we finally obtain modifications of the Friedmann equations depending on an arbitrary 
function of the scale factor. Definitely, the capabilities of such cosmological 
constructions are huge.

We considered two basic ansatzes for the aforementioned modification, namely a power law 
and an exponential one. We mention that we did not consider an explicit cosmological 
constant, since we desired to investigate the pure effects of the new terms.  In order to 
bypass the complexity of the equations, we applied the dynamical systems method, which 
allows to reveal the global behavior of time asymmetric cosmology, independently of the 
details of the evolution and the specific initial conditions. In particular, we extracted 
the critical points of the scenario  and we examined which of them are stable and thus 
they can be the late-time state of the universe, calculating also the corresponding 
observables, such as the various density parameters and the deceleration parameter.

For the power-law ansatz  we found that the universe can result in a dark-energy 
dominated, accelerating universe, where the dark-energy equation-of-state parameter 
$w_{DE}$ can lie either in the quintessence or in the phantom regime, or even behave as 
an effective cosmological constant giving rise to a de Sitter universe. Moreover, by 
suitably choosing the model parameter, one can obtain a $w_{DE}$ in agreement with 
observations.
 
For the exponential ansatz  we showed that for positive exponential coefficient at 
late times the universe is attracted by a dark-energy  dominated universe, in which 
$w_{DE}$ lies in the phantom regime, resulting finally to a finite-time Big-Rip 
singularity (due to the exponential increase of the novel terms). On the other hand, for 
negative exponential coefficient the universe results to a dark-matter dominated universe 
(due to the exponential decrease of the novel terms comparing to the matter sector), 
which is not favored by observations. Nevertheless, one could improve this behavior by 
the addition of an explicit cosmological constant, in which case he could get the correct 
thermal history, namely the succession of matter and dark-energy eras. Finally, note that 
in the case of closed curvature, the universe may experience a cosmological bounce or 
turnaround, or even cyclic behavior.
 
Concerning phenomenology, we should mention that in the scenario at hand the left 
handed neutrinos propagate differently than the photons \cite{Cortes:2015ola}, since the 
latter propagate according to the usual connection of the spacetime metric, while the 
former propagate according to the Ashtekar connection and geometry. Hence, if one desires 
to be in agreement with observations, for instance with the data from SN1987A supernova 
which show that massless neutrinos propagate similarly to photons with an error 
less than $10^{-9}$ \cite{Longo:1987ub,Hirata:1987hu,Bionta:1987qt}, then he should 
impose the new time-asymmetric modifications to be small, as expected. Interestingly 
enough, even if one considers the extreme realization of the above requirement, namely to 
assume that the new terms tend asymptotically to zero (instead 
of being increasing) as the universe expands, one can 
still have significant effects at large scales, that can radically alter the universe 
behavior (for instance in the power-law modification with  $f(V)=g_1V^{m}$ and $p=3m+1$, 
for the parameter window $0>m>-1/3$ one has an asymptotically vanishing modification term 
which is nevertheless able to drive late-time acceleration (since $1>p>0$) since its 
tends to zero as $a^{(2p-2)}$ i.e slower than the matter and curvature contributions in 
the Friedmann equation). Hence, one can 
easily pass all the cosmological tests, and definitely all the Solar System ones. An 
interesting study would be to examine the bounce realization, since in such a case one 
would expect the time asymmetry to lead to distinguishable signatures on 
observations, especially having in mind the different behavior of the spacetime and 
Ashtekar related quantities. Additionally, an important and necessary 
investigation would be to examine the cosmological 
perturbations and their relation to various observables, either at early, inflationary 
times, or at late epochs. Since both these studies lie beyond the scope of the present 
work they are left for future projects.

In summary, the cosmological application of time asymmetric extensions of 
general relativity  has many capabilities and thus it can be a good candidate for the 
description of the universe, that is worthy to be  studied further.

\begin{acknowledgments}
The authors would like to thank Lee Smolin and Radouane Gannouji for useful comments.
GL was supported by  Comisi\'on Nacional de Ciencias y Tecnolog\'ia 
through Proyecto FONDECYT de Postdoctorado 2014  grant  3140244. The research of ENS is 
implemented within the 
framework of the
Operational Program ``Education and Lifelong Learning'' (Actions
Beneficiary: General Secretariat for Research and Technology), and
is co-financed by the European Social Fund (ESF) and the Greek State.
\end{acknowledgments}

\begin{appendix}

\section{Stability of the nonhyperbolic critical points of  Model II:  $f(V)=g_2 
e^{\lambda V}$} 

In this Appendix we investigate the stability of the nonhyperbolic critical points that 
appear in the analysis of Model II in subsection \ref{ModIIanalysis}, using the center 
manifold method \cite{wiggins}, since in this case the simple linear analysis is not 
adequate.

\subsection{Zero or negative curvature  and $\lambda>0$}
\label{App1}

In the case of zero or negative curvature  and $\lambda>0$, we extract two isolated 
nonhyperbolic critical points, and a curve of nonhyperbolic critical points, displayed in 
Table \ref{Tab3}. Since point $Q_4$ and the curve $Q_5$ have at least one unstable 
eigen-direction they will definitely be non-stable (i.e. saddle or unstable), and 
hence we do not need to perform 
the center manifold analysis, since in this work we are interested in the stable 
late-time 
solutions. Thus, we restrict our analysis in the case of $Q_6$.

We introduce the new variables 
\begin{align}
\epsilon=1-T, \ \
x=\Omega_k,\ \
y=1-\Omega_{DE},
\end{align} in order to translate $Q_6$ to the origin, and thus we obtain the system 
 \begin{subequations}
 	\begin{align}
 	&\frac{d\epsilon}{d\bar{\eta}}=3 (\epsilon -1) \epsilon ^2,\\
 	&\frac{d x}{d\bar{\eta}}=-x \left\{\epsilon  (3 w_m x+x-4)-3 y\left[(w_m-1) \epsilon 
+2\right]+6\right\},\\
 	&\frac{d y}{d\bar{\eta}}=(1-y) \left\{(3 w_m+1) x \epsilon -3 y\left[(w_m-1) \epsilon 
+2\right]\right\},
 	\end{align}
 \end{subequations}
 where the local center manifold of the origin $(\epsilon, x,y)=(0,0,0)$ is tangent to 
the $\epsilon$-axis. Hence, it can be written locally as the graph
 \begin{align}
 \{(\epsilon,x,y): x=h_1(\epsilon), y=h_2(\epsilon), h_1(0)=0, h_2(0)=0, h_1'(0)=0, 
h_2'(0)=0, |\epsilon|<\delta \},
 \end{align}
 where $\delta$ is a suitably small number. 
 The functions $h_1$ and $h_2$ must satisfy the quasilinear system of differential 
equations
 \begin{subequations}
 	\label{center1}
 	\begin{align}
&3 (\epsilon -1) \epsilon ^2 h_1'(\epsilon )+h_1(\epsilon ) \left\{\epsilon  \left[3 w_m 
h_1(\epsilon 
)+h_1(\epsilon )-4\right]-3 	h_2(\epsilon ) \left[(w_m-1) \epsilon 
+2\right]+6\right\}=0,\\
& [1-h_2(\epsilon )] h_1(\epsilon ) (3 w_m  +  1)\epsilon +3
 	h_2(\epsilon ) \left[(w_m-1) \epsilon +2)\right]-3 (\epsilon -1) \epsilon ^2 
h_2'(\epsilon 
)=0.
 	\end{align}
 \end{subequations}
  This system admits the following solutions:
\begin{enumerate}
\item the point:
 \begin{align}
 &h_1(\epsilon)=0,\\
 & h_2(\epsilon)=0,
\end{align}
\item
the 1-parameter solution:
 \begin{align}
 & h_1(\epsilon)=0,\\
 & h_2(\epsilon)= \left\{\begin{array}{cc}
 \frac{\epsilon ^{w_m+1}}{e^{c_1+\frac{2}{\epsilon }}
 	(1-\epsilon )^{w_m+1}+\epsilon ^{w_m+1}}, &\ \ \epsilon\neq 0\\
 0, &\ \ \epsilon=0
 \end{array}\right.,
 \end{align}
\item  the 2-parameter solution: 
 \begin{align}
&h_1(\epsilon)=\left\{\begin{array}{cc}\frac{e^{c_2} \epsilon ^{2/3} (1-\epsilon
	)^{w_m+\frac{1}{3}}}{e^{c_2} \epsilon ^{2/3} (1-\epsilon
	)^{w_m+\frac{1}{3}}+c_1 e^{2/\epsilon } (1-\epsilon
	)^{w_m+1}+\epsilon ^{w_m+1}},& \ \  \epsilon\neq 0\\
 0,\ \  & \ \, \epsilon=0\end{array}\right.,\\
& h_2(\epsilon)=\left\{\begin{array}{cc}\frac{e^{c_2} \epsilon
	^{2/3} (1-\epsilon )^{w_m+\frac{1}{3}}+\epsilon
	^{w_m+1}}{e^{c_2} \epsilon ^{2/3} (1-\epsilon
	)^{w_m+\frac{1}{3}}+c_1 e^{2/\epsilon } (1-\epsilon
	)^{w_m+1}+\epsilon ^{w_m+1}},& \ \  \epsilon\neq 0\\
0, & \ \ \epsilon=0\end{array}\right.. 
 \end{align}
 \end{enumerate}
These three classes of solutions satisfy the smoothness conditions required  in order to 
obtain the center manifold of the origin (note that the expression for the center 
manifold is not unique).
Thus, we conclude that the evolution on the center manifold is 
given by the equation 
 \begin{align}
\frac{d\epsilon}{d\bar{\eta}}=-3 (1-\epsilon) \epsilon ^2,
 \end{align}
which admits the solution 
 \begin{align}
 \bar{\eta} =c_1+\frac{1}{3} \left[\frac{1}{\epsilon }+2 \tanh ^{-1}(1-2
 \epsilon )\right]=c_1+\frac{1}{3 \epsilon }-\frac{\log (\epsilon )}{3}-\frac{\epsilon
 }{3}+O\left(\epsilon ^2\right),
 \end{align}
and therefore by inverting the above expression we find $\epsilon(\bar{\eta})$. It is 
easy to 
see that $\epsilon \rightarrow 0$ as $\bar{\eta}\rightarrow \infty$ and that $\epsilon 
\rightarrow 1$ as $\bar{\eta}
\rightarrow -\infty$. Hence, we deduce that the center manifold of $Q_6$ is stable 
\cite{wiggins}.

\subsection{Zero or negative curvature and $\lambda<0$}
\label{App2}

In the case of zero or negative curvature  and $\lambda<0$, we extract two isolated 
nonhyperbolic critical points, and a curve of nonhyperbolic critical points, which are 
presented in Table \ref{Tab3b}. Since point $Q_{12}$ has at least two unstable 
eigen-directions it will definitely be non-stable, and hence we do not investigate it 
further.

In order to examine the stability of $Q_{10 }$ using the center manifold theorem we 
introduce the variables 
\begin{align}
\epsilon=1-{T_{1}},\ \ u=\Omega_{DE},\ \ v=\Omega_k,
\end{align}
with evolution equations given by 
\begin{subequations}
	\label{center1a} 
	\begin{align}
	& \frac{d\epsilon}{d\check{\eta}} = 3 (\epsilon -1) \epsilon ^2,\\
	& \frac{d u}{d\check{\eta}}=-u \left\{\epsilon  \left[3 w_m (u+v-1)+9 
u+v\right]-6 u-9 \epsilon 
+6\right\},\\
	& \frac{d v}{d\check{\eta}}=-v \left\{3 u \left[(w_m+3) \epsilon -2\right]+(v-1) 
(3 w_m+1) 
\epsilon\right\}.
	\end{align}
\end{subequations}
The center subspace of the origin of \eqref{center1a} is spanned by the vectors 
$(1,0,0)$ and $(0,1,0)$, which implies that the local center manifold of the origin can 
be written locally as the graph
$\{(\epsilon, u, v): v=h(\epsilon, u), h(0,0)=0, {\mathbf{Dh}}(0,0)={\mathbf{0}}, 
||(\epsilon,u)||<\delta\}
$, where $ {\mathbf{Dh}}$ is the matrix of derivatives,  $\delta$ is a suitably small 
constant, and 
  $h(\epsilon,u)$ satisfies the quasilinear partial differential equation
  \begin{align}
 & u \frac{\partial h}{\partial u} \left\{(3 w_m  + )\epsilon
 h+3 (u-1) \left[(w_m+3) \epsilon -2\right]\right\}-3 (\epsilon
 -1) \epsilon ^2 \frac{\partial h}{\partial \epsilon}
 \nonumber
 \\ & \qquad\qquad -h \left\{(3
 w_m+1) \epsilon  (h-1)+3 u \left[(w_m+3)
 \epsilon -2\right] \right\}=0.
 \label{quasilinear00}
 \end{align}
 Assuming that 
$ h(\epsilon, u)=u f(\epsilon)$ and $\lim_{\epsilon\rightarrow 0} f(\epsilon)= 
 \lim_{\epsilon\rightarrow 0}
 f'(\epsilon)=0$, and substituting in \eqref{quasilinear00}, we obtain
 \begin{align}
 3 (\epsilon -1) \epsilon ^2 f'(\epsilon )+(8
 \epsilon -6) f(\epsilon )=0,
 \end{align}
 which has the general solution 
 \begin{align}
 f(\epsilon)=\frac{c_1 e^{2/\epsilon } \epsilon ^{2/3}}{(1-\epsilon )^{2/3}},
 \end{align}
and the trivial solution $f(u)=0$. However, the general solution leads to   
$\lim_{\epsilon\rightarrow 0} 
f(\epsilon)=\text{sgn}(c_1) \infty, \lim_{\epsilon\rightarrow 0} 
f(\epsilon)=-\text{sgn}(c_1) \infty$, and hence it does not satisfy 
the imposed limits. Thus, the only accepted solution is the trivial one, which implies 
$h(\epsilon,u)\equiv 0$.
Hence, for this case the dynamics on the center manifold is governed by 
\begin{subequations}
	\begin{align}
	&\frac{d\epsilon}{d\check{\eta}} = 3 (\epsilon -1) \epsilon ^2,\\
	&\frac{d u}{d\check{\eta}}=-3 (u-1) u \left[(w_m+3) \epsilon -2\right].
	\end{align}
\end{subequations}
Eliminating time and integrating out we finally acquire
\begin{equation}
u(\epsilon)=\frac{1}{e^{c_1+\frac{2}{\epsilon }} \epsilon ^{w_m+1}
	(1-\epsilon )^{-w_m-1}+1},
\end{equation}
which satisfies $u\rightarrow 0$ as $\epsilon\rightarrow 0$. This feature implies that 
$Q_{10}$ attracts the orbits contained in its center manifold (that is the 2D set 
$T$-$\Omega_{DE}$), and thus this nonhyperbolic point behaves as stable.

In order to examine the stability of the curve of critical points $Q_{11}$ (with 
$\Omega_{kc}\in (0,1]$), using the 
center manifold theorem, we introduce the variables 
\begin{align}
\epsilon=1-{T_{1}}, \ \ u=\Omega_{k c} (1-\Omega_{DE})-\Omega_{k},\ \ v= \Omega_{DE},
\end{align}
which satisfy the evolution equations
\begin{subequations}
	\label{center2xxxxx}
	\begin{align}
	&\frac{d \epsilon}{d\check{\eta}}=-3 (1-\epsilon) \epsilon ^2,\\
	&\frac{d u}{d\check{\eta}}=\epsilon  \left\{u^2 (3
	w_m+1)+u \left[-3 v (w_m+3)+3
	w_m+1\right]\right\}
	\nonumber\\
	& \ \ \ \ \ \ \ 
	+\Omega_{k c} \epsilon  \left\{u \left[(v-2) (3
	w_m+1) \right]+(1-\Omega_{k c})(v-1) (3 w_m+1)\right\}+6
	u v,\\
	&\frac{d v}{d\check{\eta}}=\epsilon  \left[u v (3 w_m+1)
	 +3 v(1-v) (w_m+3)\right]
	 \nonumber\\ 
	 &\ \ \ \ \ \ \  +\Omega_{k c}
	\epsilon  \left[v(v-1) (3 w_m+1) \right]+6 v(v-1).
	\end{align}
\end{subequations}
Since the center subspace of the origin of \eqref{center2xxxxx} is spanned by the vectors 
$(1,0,0)$ and $(0,1,0)$, we deduce that the local center manifold of the origin can be 
written locally as the graph
$
\{(\epsilon, u, v): v=h(\epsilon, u), h(0,0)=0, {\mathbf{Dh}}(0,0)={\mathbf{0}}, 
||(\epsilon,u)||<\delta\}$, with $\delta$ a suitably small constant, and where
the function $h(\epsilon,u)$ that defines the center manifold must satisfy the 
quasilinear partial differential equation
\begin{align}
\label{hcenterxxxxx}
&\frac{\partial h}{\partial u} \Bigg\{h \Big\{(3 w_m+1)
(\Omega_{k c}-1) \Omega_{k c} \epsilon -u \left\{\epsilon 
\left[3 w_m (\Omega_{k c}-1)+\Omega_{k c}-9\right]+6\right\}\Big\} 
\nonumber \\ & \  \  \  \  \  \  \  \  
\ - \epsilon(3
w_m+1)  (u-\Omega_{k c}) (u-\Omega_{k c}+1)\Bigg\}
\nonumber \\ & -3 (\epsilon 
-1) \epsilon ^2\frac{\partial h}{\partial \epsilon}+\left\{\epsilon  \left[3 w_m 
(\Omega_{k c}-1)+\Omega_{k c}
-9\right]+6\right\} h^2
\nonumber \\ & +h
\left\{\epsilon  \left[(u-\Omega_{k c})( 3w_m+1)  +3(
w_m+3)\right]-6\right\}=0.
\end{align}
For $\Omega_{k c}\neq 1, w_m\neq -1/3$ the above equation should be integrated 
numerically. 

We will proceed using Taylor expansion. In particular, the solution $h(\epsilon,u)$ must 
satisfy the conditions $h(0,0)=0, 
{\mathbf{Dh}}(0,0)={\mathbf{0}}$,  that is it must be at least of second order in the 
variables $\epsilon$ and $u$. Hence, we assume that $h(\epsilon, u)= a_{11} \epsilon^2+ 
a_{12} \epsilon u +a_{22} u^2+\mathcal{O}(3)$, where $\mathcal{O}(3)$ denotes terms of 
third order on the vector norm, i.e. terms like $\epsilon^2 u, \epsilon u^2, \epsilon^3, 
u^3$. These terms and higher-order terms neglected in the approximation scheme. 
Substituting back this expression for $h$, neglecting third-order terms, comparing terms 
of the same power, equating to zero the coefficients, and assuming that $\Omega_{k c}\neq 
1, w_m\neq -1/3$, we obtain that a good approximation of the center manifold is given by 
\begin{equation}
h(\epsilon, u)= \frac{1}{6}a_{12}(1+3 w_m)(1-\Omega_{kc})\Omega_{kc} \epsilon^2+ a_{12} 
\epsilon u 
+\frac{3 a_{12}}{(1+3 w_m)(1-\Omega_{kc})\Omega_{kc}} u^2.
\end{equation} 
Therefore, 
we deduce that the dynamics on the center manifold is determined up to third order by 
\begin{subequations}
	\label{Numeric1}
	\begin{eqnarray}
	&&\frac{d \epsilon}{d \check{\eta}}=-3 \epsilon ^2\\
	&&\frac{d u}{d \check{\eta}}=(3 w_m+1) \epsilon  \left[(\Omega_{kc}-1) 
\Omega_{kc}-u (2 \Omega_{kc}
-1)\right].
	\end{eqnarray}
\end{subequations}
The system \eqref{Numeric1} admits the general solution 
\begin{align}
& \epsilon (\check{\eta})=\frac{1}{3 \check{\eta}-c_1},\\
& u(\check{\eta})=c_2 \left(3
   \check{\eta}-c_1\right){}^{-\frac{1}{3} (3 w_m+1) (2 
\Omega_{kc}-1)}+\frac{(\Omega_{kc}-1) \Omega_{kc}}{2
   \Omega_{kc}-1}. 
\end{align}
Observe that as $\check{\eta}\rightarrow +\infty$, $\epsilon \rightarrow 0$, but $u$ 
departs from zero and becomes unbounded in the case $\Omega_{kc}\leq\frac{1}{2}$, or 
tends to 
$\frac{(\Omega_{kc}-1) \Omega_{kc}}{2   \Omega_{kc}-1}$ for $\Omega_{kc}> \frac{1}{2}$ as 
$\check{\tau}\rightarrow +\infty$, 
which is nonzero since $\Omega_{k c}\notin \{0,1\}$. Thus, the origin is unstable along 
the $u$-axis and stable along the $\epsilon$-axis. Summarizing, the line of fixed points 
$Q_{11}$ behaves as saddle, provided that $\Omega_{k c}\neq 1, w_m\neq -1/3$.  

Let us mention that the above analysis is essentially an approximation. 
Nevertheless, there is a special point of the curve  $Q_{11}$, namely 
$(\Omega_k,\Omega_{DE}, {T_{1}})=(1,0,1)$, for which the above procedure is not valid, 
that allows 
for an analytical application of 
the center manifold analysis. 
It corresponds to $\Omega_{k c}=1$ in \eqref{center2xxxxx}. Setting $\Omega_{k c}=1$ in 
\eqref{hcenterxxxxx} we obtain the simpler quasilinear partial differential equation
\begin{align}
\label{quasilinear}
&-u  \left[(6-8 \epsilon ) h+(u-1) (3
w_m+1) \epsilon \right]\frac{\partial h}{\partial u}-3 (\epsilon -1) \epsilon ^2
\frac{\partial h}{\partial \epsilon}
\nonumber
\\  
& \qquad \ \ \ \ \qquad +h \left[(6-8 \epsilon )
h+\epsilon  (3 u w_m+u+8)-6\right]=0.
\end{align} 
Given the solution $v=h(\epsilon,u)$, the dynamics on the center manifold is determined 
by 
\begin{subequations}
	\label{center2}
	\begin{align}
	&\frac{d \epsilon}{d \check{\tau}}=-3 (1-\epsilon) \epsilon ^2\\
	&\frac{d u}{d \check{\tau}}=u \left[(u-1) (3 w_m+1) \epsilon +h(\epsilon,u) (6-8 
\epsilon )\right]. 
	\end{align}
\end{subequations}
Assuming that 
$h(\epsilon, u)=u f(\epsilon)$ and $ \lim_{\epsilon\rightarrow 0} f(\epsilon)= 
\lim_{\epsilon\rightarrow 0}
 f'(\epsilon)=0$, and substituting into \eqref{quasilinear}, we obtain
\begin{align}
u \left\{f(\epsilon ) \left[(w_m+3) \epsilon -2\right]-(\epsilon -1)
\epsilon ^2 f'(\epsilon )\right\}=0,
\end{align}
which has the general solution 
\begin{align}
f(\epsilon )=\left\{\begin{array}{cc}
 c_1 e^{-2/\epsilon } (1-\epsilon )^{w_m+1}
\epsilon ^{-w_m-1}, & \epsilon\neq 0\\
 0, & \epsilon=0
\end{array}\right., 
\end{align}
which indeed satisfies the imposed limits. 
Hence, the dynamics on the center manifold is governed by the evolution equations
\begin{subequations}
	\label{Syst_A}
	\begin{align}
	&\frac{d \epsilon}{d \check{\tau}}=-3 (1-\epsilon) \epsilon ^2\\
	&\frac{d u}{d \check{\tau}}=u \left[c_1 u e^{-2/\epsilon } (6-8 \epsilon ) 
(1-\epsilon
	)^{w_m+1} \epsilon ^{-w_m-1}+(u-1) (3 w_m+1)
	\epsilon \right].
	\end{align}
\end{subequations}
Eliminating the time variable we find that the system \eqref{Syst_A} can be expressed as 
\begin{align}
3 (1-\epsilon) \epsilon	\frac{d u(\epsilon)}{d \epsilon}=(3 w_m+1) u(\epsilon) + \mu 
(\epsilon )u(\epsilon)^2
\end{align}
with 
$\mu(\epsilon)=2 c_1 e^{-2/\epsilon } (4 \epsilon -3) (1-\epsilon )^{w_m+1}
\epsilon ^{-w_m-2}-3 w_m-1$,
which admits the quadrature
\begin{align}
u\left( \epsilon \right) =\frac{(1-\epsilon )^{-w_m-\frac{1}{3}} \epsilon
	^{w_m+\frac{1}{3}}}{c_2-\int  \frac{1}{3}
	\mu (\epsilon) (1-\epsilon)^{-w_m-\frac{4}{3}}
	\epsilon^{w_m-\frac{2}{3}} \, d\epsilon}.
\end{align}
Since  $\mu(\epsilon)\rightarrow -3 w_m-1$ as $\epsilon\rightarrow 0$, we can  
integrate the above quadrature in the approximation $\epsilon\rightarrow 0$, obtaining 
\begin{align}
u(\epsilon)\approx \frac{1}{c_2 (1-\epsilon )^{w_m+\frac{1}{3}} \epsilon
	^{-w_m-\frac{1}{3}}+1},
\end{align}
which tends to zero as $\epsilon\rightarrow 0$, for $w_m >-\frac{1}{3}$.
Hence, we deduce that  the center manifold associated to the point $(1,0,1)$ is stable. 
Indeed, this behavior is the typical one for $w_m >-\frac{1}{3}$, as can be verified 
by Fig. \ref{Fig3}.

 \subsection{Positive curvature  and $\lambda>0$}
\label{App3}
In the case of positive curvature  and $\lambda>0$, we extract two isolated nonhyperbolic 
critical points, and a curve of nonhyperbolic critical points corresponding to expansion, 
which are presented in Table \ref{Tab4}. Each of the above points/curve in Table 
\ref{Tab4} has a partner through the symmetry \eqref{discrete_symm_II}, which represents 
a contracting cosmology, and are displayed in Table \ref{Tab4b}. Amongst them in this 
Appendix we analyze only the nonhyperbolic fixed points that might be late-time 
attractors (for instance points like $Q_{15}$ and the curve of critical points $Q_{16}$ 
that have at least one unstable eigen-direction will definitely be either unstable or 
saddle and thus we do not investigate them further). These are the expanding solution 
$Q_{17}$ and the contracting solutions $R_{15}$ and $R_{16}$. 
We remind that the points $Q_i$ and their contracting partners points $R_i$ through the 
symmetry \eqref{discrete_symm_II}, exhibit opposite dynamical behaviors, and thus from 
the following analysis we also obtain information for the contracting solution $R_{17}$
and the expanding ones $Q_{15}$ and $Q_{16}$.


In order to examine the stability of the contracting solution $R_{15}$ we introduce the 
variables 
\begin{align}
\epsilon=1-T,\ \ x=1+Q_0,\ \ y=\Theta_{DE},
\end{align}
and therefore the autonomous system \eqref{systII4} becomes 
\begin{subequations}
\begin{align}
\label{R15}
&\frac{d \epsilon}{d \bar{\tau}}=-3 (1-x) (\epsilon -1) \epsilon
   ^2,\\
&\frac{d x}{d \bar{\tau}}=	-\frac{1}{2} (x-2) x (\epsilon  (3
   w_m (y-1)-3 y-1)+6 y),\\
&\frac{d y}{d \bar{\tau}}=-3 (1-x) (1-y) y
   [(w_m-1) \epsilon +2].
\end{align}
\end{subequations}
The center manifold of the origin of \eqref{R15} is spanned by the vectors 
$(1,0,0)$ and $(0,1,0)$, which implies that the local center manifold of the origin can 
be written locally as the graph
$
\{(\epsilon, x,y): y=h(\epsilon, x), h(0,0)=0, {\mathbf{Dh}}(0,0)={\mathbf{0}}, 
||(\epsilon,u)||<\delta\}$,
with  $\delta$   a suitably small constant and   $ {\mathbf{Dh}}$   the matrix of 
derivatives. The function $h(\epsilon,x)$ satisfies the quasilinear partial differential 
equation 
\begin{align}
\label{centerR15}
  &\!\!\!\!\!\!\!\!\!\!\!\!\!\!\!\!\!\!\!\!\!\! \frac{1}{2} (x-2) x \frac{\partial 
h}{\partial x} \Big[3
   {(w_m-1) \epsilon +2} h-(3 w_m+1) \epsilon \Big]-3 (x-1)
   (\epsilon -1) \epsilon ^2 \frac{\partial h}{\partial \epsilon}\nonumber\\
	&\ \ -3 (x-1) [(w_m-1) \epsilon +2]
   (h-1) h=0.
\end{align}
The equation \eqref{centerR15} admits the solutions: 
\begin{enumerate}
\item The trivial solution $h(\epsilon,x)=0$,
 \item the one-parameter solution 
$h(\epsilon,x)=\left\{
\begin{array}{cc}
\frac{e^{2/\epsilon } (\epsilon -1)
   (1-\epsilon )^{w_m}}{-e^{c_1}
   \epsilon ^{w_m+1}-e^{2/\epsilon }
   (1-\epsilon )^{w_m}+e^{2/\epsilon
   } \epsilon  (1-\epsilon )^{w_m}} &  \epsilon\neq 0\\
1 & \epsilon=0
\end{array}\right.$.
\end{enumerate}
Only the trivial solution satisfies the conditions $ h(0,0)=0, 
{\mathbf{Dh}}(0,0)={\mathbf{0}}$. 
Henceforth, the dynamics on the center manifold is governed by 
\begin{subequations}
\begin{align}
&\frac{d \epsilon}{d \bar{\tau}}=3 (1-x) (1-\epsilon) \epsilon^2,\\
&\frac{d x}{d \bar{\tau}}=\frac{1}{2} (3 w_m+1) (x-2) x
   \epsilon. 
\end{align}
\end{subequations}
Eliminating the time variable, $\bar{\tau}$, and using the chain rule for derivatives we 
find that the orbits on the invariant manifold satisfy 
\begin{equation}
x'(\epsilon )=\frac{(3 w_m+1)
   (x(\epsilon )-2) x(\epsilon )}{6
   (\epsilon -1) \epsilon  (x(\epsilon
   )-1)}, 
\end{equation}
which admits the general solutions
\begin{align}
x(\epsilon)=1\pm \epsilon ^{-w_m-\frac{1}{3}}
   \sqrt{\epsilon ^{w_m+\frac{1}{3}}
   \left(\epsilon
   ^{w_m+\frac{1}{3}}-e^{2 c_1}
   (1-\epsilon
   )^{w_m+\frac{1}{3}}\right)}.
\end{align}
None of these solutions satisfy the condition $x(0)=0$, indeed $x$ becomes infinity  as 
$\epsilon \rightarrow 0$. Thus, any solution starting with $\epsilon\neq 0$ and $x\neq 0$ 
departs from the origin along the $x$-direction, which implies that $R_{15}$ behaves as a 
saddle.

In order to examine the stability of the contracting solution $R_{16}$ we introduce the 
variables 
 \begin{equation}
   u_1=Q_0-\frac{\left({Q_{0 c}}^2-
   1\right) \Theta_{DE}}{2
   {Q_{0 c}}}+{Q_{0 c}},
	\; u_2=\frac{1}{2}
   \left({Q_{0 c}}^2-1\right) (1-T) (3
   w_m+1),
\; v=\Theta_{DE},
\end{equation}	
where $Q_{0 c}$ is a constant $Q_{0 c}\in (0,1)$, which satisfy the equations
\begin{subequations}
\label{evolR16}
\begin{align}
&\frac{d u_1}{d \bar{\tau}}=\frac{3 (v-1) v \left[{Q_{0 c}}^2
   (v-2)+2 {Q_{0 c}} {u_1}-v\right]
   \left[{w_m} \left(3
   {Q_{0 c}}^2+{u_2}-3\right)+{Q_{0 c}}^2-{u_2}-1\right]}{2
   {Q_{0 c}}^2 (3
   {w_m}+1)}
   \nonumber\\
   &
   -\frac{1}{2}
   \left[\left(-\frac{{Q_{0 c}}
   v}{2}+\frac{v}{2
   {Q_{0 c}}}+{Q_{0 c}}-{u_1}\right)
   ^2-1\right] \left[\frac{2 {u_2} (3
   v ({w_m}-1)-3
   {w_m}-1)}{\left({Q_{0 c}}^2-1\right) (3 {w_m}+1)}+6 v\right],\\
	&\frac{d u_2}{d \bar{\tau}}= \frac{3
   {u_2}^2 \left[{Q_{0 c}}^2 (v-2)+2
   {Q_{0 c}} {u_1}-v\right]
   \left[{Q_{0 c}}^2 (-(3 {w_m}+1))+2
   {u_2}+3
   {w_m}+1\right]}{{Q_{0 c}}
   \left({Q_{0 c}}^2-1\right)^2 (3
   {w_m}+1)^2},\\
	& \frac{d v}{d \bar{\tau}}=3 (v-1) v
   \left[-\frac{{Q_{0 c}}
   v}{2}+\frac{v}{2
   {Q_{0 c}}}+{Q_{0 c}}-{u_1}\right]
   \left[\frac{2 {u_2}
   ({w_m}-1)}{\left({Q_{0 c}}^2-1\right) (3 {w_m}+1)}+2\right].
	\end{align}
	\end{subequations}
 Note that by definition $u_2\leq 0$. 

The center subspace of the origin of \eqref{evolR16} is spanned by the vectors 
$(1,0,0)$ and $(0,1,0)$, which implies that the local center manifold of the origin can 
be written locally as the graph
$
\{(u_1,u_2, v): v=h(\epsilon, u), h(0,0)=0, {\mathbf{Dh}}(0,0)={\mathbf{0}}, 
||(u_1,u_2)||<\delta\}$,
with  $\delta$   a suitably small constant and   $ {\mathbf{Dh}}$   the matrix of 
derivatives. The function $h(u_1, u_2)$ satisfies the quasilinear partial differential 
equation
\begin{align}
\label{R16eqh}
&
\!\!\!\!\!
\frac{3 {u_2}^2 \left[3
   \left({Q_{0 c}}^2-1\right)
   {w_m}+{Q_{0 c}}^2-2
   {u_2}-1\right]
     \left[\left({Q_{0 c}}^2-1\right)
   {h}+2 {Q_{0 c}}
   ({u_1}-{Q_{0 c}})\right]}{{Q_{0 c}} \left({Q_{0 c}}^2-1\right)^2 (3
   {w_m}+1)^2} \frac{\partial h}{\partial u_2}
   \nonumber \\
   &
   +\frac{1}{2}
   \frac{\partial h}{\partial u_1}
   \left\{\left\{\left[\frac{\left({Q_{0 c}}^2-1\right) {h}}{2
   {Q_{0 c}}}-{Q_{0 c}}+{u_1}\right]
   ^2-1\right\} \left\{\frac{2 {u_2} [3
   ({w_m}-1) {h}-3
   {w_m}-1]}{\left({Q_{0 c}}^2-1\right) (3 {w_m}+1)}+6
   {h}\right\}\right.
   \nonumber
   \\ 
   &\left. 
   -\frac{3
   \left[{w_m} \left(3
   {Q_{0 c}}^2+{u_2}-3\right)+{Q_{0 c}}^2-{u_2}-1\right]
   (h-1)
   {h}
   \left[\left({Q_{0 c}}^2-1\right)
   {h}+2 {Q_{0 c}}
   ({u_1}-{Q_{0 c}})\right]}{{Q_{0 c}}^2 (3 {w_m}+1)}\right\}
   \nonumber \\ & +3
   \left[\frac{2 {u_2}
   ({w_m}-1)}{\left({Q_{0 c}}^2-1\right) (3 {w_m}+1)}+2\right]
   (h-1)
   {h}
   \left[-\frac{\left({Q_{0 c}}^2-1\right
   ) {h}}{2
   {Q_{0 c}}}+{Q_{0 c}}-{u_1}\right]=0.
\end{align} 
Assuming that $h$ is locally given by $h= a_{11} u_1^2+ a_{1 2} u_1 u_2 +a_{2 2} u_2^2+ 
\mathcal{O}(3),$ where $\mathcal{O}(3)$ denotes terms of third order in the vector norm, 
plugging back into \eqref{R16eqh}, comparing equal powers in the variables $u_1$ and 
$u_2$ and equating to zero the corresponding coefficients, we obtain up to third order 
that $a_{1 1}=-3 a_{1 2} Q_{0 c}$ and $a_{2 2}= -
\frac{a_{12}}{6 Q_{0 c}}$. That is, the graph of the center manifold is given up to third 
order by 
$v= -3 a_{1 2} Q_{0 c} u_1^2+ a_{1 2} u_1 u_2 -\frac{a_{12}}{6 Q_{0 c}} u_2^2$.  
Therefore, by neglecting the third-order terms we find that the dynamics on the center 
manifold is governed by equations
\begin{subequations}
\label{R16center}
\begin{align}
&\frac{d u_1}{d \bar{\tau}}=u_2+\frac{2 Q_{0c}
    u_1 u_2}{1-Q_{0c}^2},\\
&\frac{d u_2}{d \bar{\tau}}=	-\frac{6 Q_{0c} u_2^2}{\left(1-Q_{0c}^2\right) (3 w_m+1)}.
\end{align}
\end{subequations}
Integrating \eqref{R16center} it follows
\begin{subequations}
\begin{align}
& u_1(\bar{\tau} )= \frac{2 c_2 Q_{0c}
   \left[c_1 \left(Q_{0c}^2-1\right) (3
   w_m+1)+6 Q_{0c} \bar{\tau}
   \right]{}^{w_m+\frac{1}{3}}+Q_{0c}^
   2-1}{2 Q_{0c}},\\
& u_2(\bar{\tau} )=
   -\frac{\left(Q_{0c}^2-1\right) (3
   w_m+1)}{c_1 \left(Q_{0c}^2-1\right)
   (3 w_m+1)+6 Q_{0c} \bar{\tau} }.
	\end{align}
\end{subequations}
Taking the limit as $\bar{\tau}\rightarrow \infty$ in the above expressions we obtain 
$(u_1, u_2)\rightarrow (c_2 \infty ,0), c_2\neq 0, u_2\neq 0$. In the special case 
$c_2=0$ we obtain the limits 
$(u_1, u_2)\rightarrow (\frac{Q_{0c}^2-1}{2 Q_{0c}} ,0).$ In both cases the origin is 
unstable along the $u_1$-axis. Since it is stable along the $u_2$-axis, it follows that 
$R_{16}$ is a saddle.

Finally, let us examine the stability of $Q_{17}$ using the center manifold theorem. We 
introduce the variables
\begin{align}
\epsilon=1-T,\ \ x=1-Q_0,\ \ y=1-\Theta_{DE},
\end{align}
and therefore the autonomous system \eqref{systII4} is equivalent to the system
\begin{subequations}
	\begin{align}
	&\frac{d \epsilon}{d \bar{\tau}}=-3 (1-x) (1-\epsilon) \epsilon ^2,\\
	&\frac{d x}{d \bar{\tau}}=-\frac{1}{2} (x-2) x \left[3
	y \left\{(w_m-1) \epsilon +2\right\}+4 \epsilon -6\right],\\
	&\frac{d y}{d \bar{\tau}}=-3 (1-x) (1-y) y
	\left[(w_m-1) \epsilon +2\right].
	\end{align}
\end{subequations}
The local center manifold of the origin $(\epsilon, x,y)=(0,0,0)$ is tangent to the 
$\epsilon$-axis.
 Thus, it can be written locally as the graph
$
\{(\epsilon,x,y): x=h_1(\epsilon), y=h_2(\epsilon), h_1(0)=0, h_2(0)=0, h_1'(0)=0, 
h_2'(0)=0, |\epsilon|<\delta \},
$
with $\delta$ a suitably small number. 
The functions $h_1$ and $h_2$ must satisfy the quasilinear system of differential 
equations
\begin{subequations}
\label{system111}
	\begin{align}
	&[h_1(\epsilon )-2]  h_1(\epsilon ) [3
	h_2(\epsilon ) ((w_m-1) \epsilon +2)+4 \epsilon
	-6]-6 (\epsilon -1) \epsilon ^2 ( h_1(\epsilon )-1)
	 h_1'(\epsilon )=0,
\\
	& [ h_1(\epsilon )-1]
	\left[(h_2(\epsilon )-1)  h_2(\epsilon )
	((w_m-1) \epsilon +2)-(\epsilon -1) \epsilon ^2
	 h_2'(\epsilon )\right]=0,
	\end{align}
\end{subequations}
which admits the general solution satisfying the conditions $h_1(0)=0, h_2(0)=0, 
h_1'(0)=0, 
h_2'(0)=0$, namely:
\begin{subequations}
	\label{CENTER_A}
	\begin{align}
		& h_1(\epsilon )= 	\left\{\begin{array}{cc}
		1-\frac{\sqrt{ e^{2 c_2} \epsilon ^{2/3} (1-\epsilon
				)^{w_m+\frac{1}{3}}+e^{c_1+\frac{2}{\epsilon }}
				(1-\epsilon )^{w_m+1}+\epsilon
				^{w_m+1}}  }{\sqrt{ e^{c_1+\frac{2}{\epsilon }}
				(1-\epsilon )^{w_m+1}+\epsilon ^{w_m+1}}}, & \epsilon\neq 
0\\
		0, & \epsilon=0
		\end{array}
		\right.,\\
		& h_2(\epsilon )=
	\left\{\begin{array}{cc}
	\frac{\epsilon ^{w_m+1}}{e^{c_1+\frac{2}{\epsilon }}
		(1-\epsilon )^{w_m+1}+\epsilon ^{w_m+1}}, & \epsilon\neq 0\\
	0, & \epsilon=0
	       \end{array}
	\right.,
	\end{align}
\end{subequations}
where $c_1$ and $c_2$ are integration constants, as well as the trivial solution 
$h_1(\epsilon)=0, h_2(\epsilon)=0$. Note that the expression for the center manifold of 
the origin is not necessarily unique.

For the expression of the center manifold of the origin given by \eqref{CENTER_A}, the 
dynamics on 
it is given by  
\begin{align}\label{CENTER_AA}
	\frac{d \epsilon}{d \bar{\tau}}=-3 f(\epsilon)(1-\epsilon) \epsilon ^2,
\end{align}
where 
\begin{equation}
f(\epsilon)=\frac{\sqrt{ e^{2 c_2} \epsilon ^{2/3} (1-\epsilon
		)^{w_m+\frac{1}{3}}+e^{c_1+\frac{2}{\epsilon }}
		(1-\epsilon )^{w_m+1}+\epsilon
		^{w_m+1}}  }{\sqrt{ e^{c_1+\frac{2}{\epsilon }}
		(1-\epsilon )^{w_m+1}+\epsilon ^{w_m+1}}}.
\end{equation}
Since $f(\epsilon)>0$, the flow of \eqref{CENTER_AA} is equivalent to the flow of 
\begin{align}\label{CENTER_AAA}
\frac{d \epsilon}{d \xi}=-3 (1-\epsilon) \epsilon ^2,
\end{align}
where we have introduced a time rescaling. 
The general solution of \eqref{CENTER_AAA} reads
\begin{align}
\xi=c_3+\frac{1}{3} \left[\frac{1}{\epsilon }+2 \tanh ^{-1}(1-2
\epsilon )\right]=c_3+\frac{1}{3 \epsilon }-\frac{\log (\epsilon )}{3}-\frac{\epsilon
}{3}+O\left(\epsilon ^2\right).
\end{align}
Since $\epsilon \rightarrow 0$ as $\xi\rightarrow \infty$, the center manifold of 
$Q_{17}$ is stable, and it corresponds to the late-time attractor. Additionally  
$\epsilon \rightarrow 1$ as $\xi\rightarrow -\infty$. Note that the relation with the 
original time variable is obtained through the quadrature  
\begin{equation}
\bar{\tau}=\int \frac{\xi'(\epsilon) d\epsilon}{f(\epsilon)}=\frac{1}{3} \int 
\frac{\sqrt{e^{c_1+\frac{2}{\epsilon }}
		(1-\epsilon )^{w_m+1}+\epsilon ^{w_m+1}}}{(\epsilon
	-1) \epsilon ^2 \sqrt{e^{2 c_2} \epsilon ^{2/3} (1-\epsilon
		)^{w_m+\frac{1}{3}}+e^{c_1+\frac{2}{\epsilon }}
 (1-\epsilon )^{w_m+1}+\epsilon ^{w_m+1}}} \,
d\epsilon.
\end{equation}

If the center manifold is given by the trivial solution $h_1(\epsilon)=0, 
h_2(\epsilon)=0$ 
we 
deduce that 
the evolution on it is dictated by  
\begin{align}
	\frac{d \epsilon}{d\bar{\tau}}=-3 (1-\epsilon) \epsilon ^2,
\end{align}
which admits the solution 
\begin{align}
\bar{\tau} =c_1+\frac{1}{3} \left[\frac{1}{\epsilon }+2 \tanh ^{-1}(1-2
\epsilon )\right]=c_1+\frac{1}{3 \epsilon }-\frac{\log (\epsilon )}{3}-\frac{\epsilon
}{3}+O\left(\epsilon ^2\right).
\end{align}
Since $\epsilon \rightarrow 0$ as $\bar{\tau}\rightarrow \infty$, the center manifold of 
$Q_{17}$ 
is 
stable, and it corresponds to the late-time attractor.

\subsection{Positive curvature  and $\lambda<0$}
\label{App4}

In the case of positive curvature  and $\lambda<0$, we extract two isolated nonhyperbolic 
critical points, and a curve of nonhyperbolic critical points, representing expanding 
solutions, which are presented in Table \ref{Tab4IIa}. Each of the above points/curve in 
Table \ref{Tab4IIa} has a partner through the symmetry \eqref{discrete_symm_IIb}, which 
represents a contracting cosmology, and are displayed in Table  \ref{Tab4IIb}. Amongst 
them we analyze only the nonhyperbolic fixed points that might be late-time  attractors 
(for instance points having at least one unstable eigen-direction, like $Q_{22}$, are 
excluded from the analysis). These are the expanding solutions $Q_{20}$ and $ Q_{21}$ 
and the contracting one $R_{22}$. We remind that the points $Q_i$ and their contracting 
partners points $R_i$ through the symmetry \eqref{discrete_symm_IIb}, exhibit opposite 
dynamical behaviors, and hence from the following analysis we also obtain information 
for the contracting solutions $R_{20}, R_{21}$ and the expanding one $Q_{22}$.

In order to calculate the center manifold of $Q_{20}=(1,0,1)$ for the system 
\eqref{systII4b}  we introduce the variables
\begin{align}
\epsilon=1-{T_{1}},\ \ u=1-Q_0, \  \ v=\Theta_{DE},
\end{align} 
which satisfy the evolution equations
\begin{subequations}
	\label{evolQ20}
	\begin{align}
	&\frac{d \epsilon}{d\check{\tau}}=-3 (1-u) (1-\epsilon) \epsilon ^2, 
\label{evolQ20a}\\
	&\frac{d u}{d\check{\tau}}=-\frac{1}{2} (2-u) u
	\left[\epsilon  (3 v (w_m+3)-3 w_m-1)-6 v\right], \label{evolQ20b}\\
	&\frac{d v}{d\check{\tau}}=3 (1-u) (1-v)
	v \left[(w_m+3) \epsilon -2\right].
\label{evolQ20c}
	\end{align}
\end{subequations}
The center subspace of the origin of \eqref{evolQ20} is spanned by the vectors 
$(1,0,0)$ and $(0,1,0)$, which implies that the local center manifold of the origin can 
be written locally as the graph
$
\{(\epsilon, u, v): v=h(\epsilon, u), h(0,0)=0, {\mathbf{Dh}}(0,0)={\mathbf{0}}, 
||(\epsilon,u)||<\delta\}$,
with  $\delta$   a suitably small constant and   $ {\mathbf{Dh}}$   the matrix of 
derivatives. The function $h(\epsilon,u)$ satisfies the quasilinear partial differential 
equation 
\begin{align}
\label{eqA59}
  &
  \!\!\!\!\!\!  \!\!\!\!\!\!  \!\!\!\!\!\!
  -\frac{1}{2} (u-2) u \frac{\partial h}{\partial u} \left[3
  \left\{(w_m+3) \epsilon -2\right\} h-(3 w_m+1)
  \epsilon\right]+3 (u-1) (\epsilon -1) \epsilon ^2 \frac{\partial h}{\partial \epsilon}
\nonumber 
\\
  &\ \ \ \ \ \ \ \ \  +3 (u-1) \left\{(w_m+3) \epsilon -2\right\} (h-1)
  h=0,
\end{align}
 which admits the formal general solution 
 \begin{align}
 F\left(\frac{1}{2} \ln \left[\frac{(u-2) u (1-\epsilon
 	)^{2/3}}{\epsilon ^{2/3} h(\epsilon ,u)}\right],\ln
 \left[\frac{e^{-2/\epsilon } (\epsilon -1) (1-\epsilon
 	)^{w_m} \epsilon ^{-w_m-1} (h(\epsilon
 	,u)-1)}{h(\epsilon ,u)}\right]\right)=0,
 \end{align}
and the trivial solution $h\equiv 0$. 
Nevertheless, in order to complete the analysis  numerical investigation is required. 
Using Taylor expansion we obtain that, up to third order in the vector norm, the solution 
of \eqref{eqA59} is the trivial solution $h\equiv 0$. Thus, the dynamics on the center 
manifold is given up to the same order by 
\begin{subequations}
\begin{align}
&\frac{d \epsilon}{d\check{\tau}}=-3 (1-u) \epsilon ^2+\mathcal{O}(3),\\
&\frac{d u}{d\check{\tau}}=u (3 w_m+1) \epsilon +\mathcal{O}(3).
\end{align}
\end{subequations}
Neglecting the error terms, and eliminating the time variable, we obtain the equation
\begin{equation}
\epsilon'(u)=\frac{3 (u-1) \epsilon (u)}{u (3 w_m+1)},
\end{equation}
which admits the solution 
\begin{equation}
\epsilon (u)=c_1 e^{\frac{3 (u-\ln (u))}{3 w_m+1}},
\end{equation}
that satisfy $\lim_{u\rightarrow 0}=c_1 \infty$, and hence it is infinity unless $c_1=0$. 
Therefore, $Q_{20}$ is a saddle for $w_m\geq 0$, as shown in Figure \ref{Fig5}. Finally, 
by symmetry, $R_{20}$ behaves as saddle too.

In order to examine the stability of the curve of critical points $Q_{21}$, using the 
center manifold theorem, we introduce the variables 
 \begin{align}
u_1= Q_{0c}-Q_0 +\frac{\left(1-Q_{0c}^2\right) \Theta_{DE}}{2 Q_{0c}},  
u_2=\frac{1}{2}\left(1-Q_{
0c}^2\right)(1-{T_{1}})(1+3 w_m), v= \Theta_{DE},
\end{align} where $Q_{0c}\in (0, 1)$ is a constant,  
which satisfy the evolution equations
\begin{subequations}
	\label{evolQ}
	\begin{align}
&\frac{d u_1}{d \check{\tau}}=\frac{3 v (v-1) 
   \left[Q_{0c}^2 (v-2)+2 Q_{0c}
   u_1-v\right] \left[w_m
   \left(3
   Q_{0c}^2+u_2-3\right)+Q_{0c}^2+3
   u_2-1\right]}{2 Q_{0c}^2 (3
   w_m+1)}
   \nonumber \\ 
   &\ +\left[\left(-\frac{Q_{0c} v}{2}+\frac{v}{2
   Q_{0c}}+Q_{0c}-u_1\right)^2-1\right] \left\{\frac{u_2
   [-3 v (w_m+3)+3
   w_m+1]}{\left(Q_{0c}^2-1\right) (3 w_m+1)}-3
   v\right\},\\
&\frac{d u_2}{d \check{\tau}}=-\frac{3 u_2^2
   \left[Q_{0c}^2 (v-2)+2 Q_{0c}
   u_1-v\right] \left[3
   \left(Q_{0c}^2-1\right)
   w_m+Q_{0c}^2+2
   u_2-1\right]}{Q_{0c}
   \left(Q_{0c}^2-1\right)^2 (3
   w_m+1)^2},\\
&\frac{d v}{d \check{\tau}}=-3 v (v-1)
   \left(-\frac{Q_{0c}
   v}{2}+\frac{v}{2
   Q_{0c}}+Q_{0c}-u_1\right)
   \left[-\frac{2 u_2
   (w_m+3)}{\left(Q_{0c}^2-1\right) (3
   w_m+1)}-2\right].
	\end{align}
\end{subequations}

The center subspace of the origin of \eqref{evolQ} is spanned by the vectors 
$(1,0,0)$ and $(0,1,0)$, and hence the local center manifold of the origin can be 
written locally as the graph
$\{(u_1, u_2, v): v=h(u_1, u_2), h(0,0)=0, {\mathbf{Dh}}(0,0)={\mathbf{0}}, 
||(u_1,u_2)||<\delta\}$,
with $\delta$ a suitably small constant, and where the function $h$ 
satisfies the quasilinear partial differential equation
\begin{align}
&
\!\!\!\!\!\!
\frac{3 u_2^2 \left[3\left(Q_{0c}^2-1\right)w_m+Q_{0c}^2+2 u_2-1\right]
      \left[\left(Q_{0c}^2-1\right) 
h+2Q_{0c}(u_1-Q_{0c})\right]}{Q_{0c}\left(Q_{0c}^2-1\right)^2
   (3w_m+1)^2} \frac{\partial h}{\partial u_2}
   \nonumber \\ 
   & -\left\{\frac{3\left[w_m 
\left(3Q_{0c}
^2+u_2-3\right)+Q_{0c}^2+3
   u_2-1\right]
   (h-1)
   h
   \left[\left(Q_{0c}^2-1\right) h+2Q_{0c}(u_1-Q_{0c})\right]}{2 Q_{0c}^2 (3w_m+1)} 
\right.
\nonumber \\ & 
+\left.
\left\{\left[\frac{\left(Q_{0c}^2-1\right) 
h}{2Q_{0c}}-Q_{0c}+u_1\right]^2-1\right\}
\left\{\frac{ u_2 \left[
-3 (w_m+3)
   h+3w_m+1\right]}{\left(Q_{0c}^2-1\right) (3w_m+1)}-3 h\right\}\right\}
   \frac{\partial 
h}{\partial u_1} 
\nonumber \\
	& 3 \left[\frac{2
   u_2(w_m+3)}{\left(Q_{0c}^2-1\right) (3w_m+1)}+2\right](h-1) h
   \left[-\frac{\left(Q_{0c}^2-1\right) h}{2Q_{0c}}+Q_{0c}-u_1\right]=0.
\end{align}
 Assuming that 
$ h(u_1, u_2)=a_{11} u_1^2 + a_{12} u_1 u_2 + a_{22} u_2^2 +\mathcal{O}(3)$ and 
evaluating 
up to 
third order,  we find a good approximation of 
the center manifold given by 
\begin{align}
h(u_1, u_2)=-3 a_{12} Q_{0c} u_1^2 + a_{12} u_1 u_2 - \frac{a_{12}}{6 Q_{0c}} u_2^2 
+\mathcal{O}(3).
\end{align}
Therefore, neglecting the $\mathcal{O}(3)$-terms,  the evolution on the center manifold 
is 
governed 
by the  equations
\begin{subequations}
\label{centeryyyyy}
\begin{align}
&\frac{d u_1}{d\check{\tau}}=u_2+\frac{2 Q_{0c}
    u_1 u_2}{1-Q_{0c}^2},\\
&\frac{d u_2}{d\check{\tau}}=	-\frac{6 Q_{0c} u_2^2}{\left(1-Q_{0c}^2\right) (3 w_m+1)},
\end{align}
\end{subequations}
where by definition, $u_2\geq 0$.

Integrating \eqref{centeryyyyy} we acquire
\begin{subequations}
\begin{align}
& u_1(\check{\tau} )= \frac{2 c_2 Q_{0c}
   \left[c_1 \left(Q_{0c}^2-1\right) (3
   w_m+1)+6 Q_{0c} \check{\tau}
   \right]{}^{w_m+\frac{1}{3}}+Q_{0c}^
   2-1}{2 Q_{0c}},\\
& u_2(\check{\tau} )=
   -\frac{\left(Q_{0c}^2-1\right) (3
   w_m+1)}{c_1 \left(Q_{0c}^2-1\right)
   (3 w_m+1)+6 Q_{0c} \check{\tau} }.
	\end{align}
\end{subequations}
Taking the limit $\check{\tau}\rightarrow \infty$ in the above expressions we obtain 
$(u_1, u_2)\rightarrow (c_2 \infty ,0), c_2\neq 0, u_2\neq 0$. In the special case where
$c_2=0$ we obtain the limits 
$(u_1, u_2)\rightarrow (\frac{Q_{0c}^2-1}{2 Q_{0c}} ,0).$ In both cases the origin is 
unstable 
along the $u_1$-axis. Since it is stable along the $u_2$-axis, it follows that $Q_{21}$ 
is 
a saddle.

Another nonhyperbolic point that can be a late-time attractor is the contracting point 
$R_{22}$, which has a 2D stable manifold. Introducing the new variables 
\begin{equation}
\epsilon=1-{T_{1}}, x=Q_0+1, y=1-\Theta_{DE},
\end{equation}
 we obtain the equivalent dynamical system
\begin{subequations}
\begin{align}
&\frac{d\epsilon}{d\check{\tau}}=3 (x-1) (\epsilon -1) \epsilon
   ^2,\\
& \frac{d x}{d\check{\tau}}=	\frac{1}{2} (x-2) x \{3 y [(w_m+3)
   \epsilon -2]-8 \epsilon +6\},\\
& \frac{d y}{d\check{\tau}}=	3 (x-1) (y-1) y
   [(w_m+3) \epsilon -2],
\end{align}
\end{subequations}
  where the local center manifold of the origin $(\epsilon, x,y)=(0,0,0)$ is tangent to 
the $\epsilon$-axis. Hence, it can be written locally as the graph
 \begin{align}
 \{(\epsilon,x,y): x=h_1(\epsilon), y=h_2(\epsilon), h_1(0)=0, h_2(0)=0, h_1'(0)=0, 
h_2'(0)=0, |\epsilon|<\delta \},
 \end{align}
 where $\delta$ is a suitably small number. 
 The functions $h_1$ and $h_2$ must satisfy the quasilinear system of differential 
equations
 \begin{subequations}
 	\label{centerR22}
 	\begin{align}
&(h_1-2)
   h_1 \left\{3 h_2
   [(w_m+3) \epsilon -2]-8 \epsilon +6\right\}-6
   (\epsilon -1) \epsilon ^2 (h_1-1) h_1'=0,\\
&	(h_1-1)
   \left\{(h_2-1)
   h_2 [(w_m+3) \epsilon
   -2]-(\epsilon -1) \epsilon ^2
   h_1\right\}=0.
 	\end{align}
 \end{subequations}
These equations admit the following classes of solutions:
\begin{enumerate}
\item The trivial solution $h_1\equiv 0, h_2\equiv 0,$
\item the one-parameter class of solutions 
\begin{subequations}
\begin{align}
&h_1=
1\pm \sqrt{1-\frac{\epsilon ^{2/3} e^{2
   \left(c_1+\frac{1}{\epsilon
   }\right)}}{(1-\epsilon )^{2/3}}},\\
& h_2=0,
\end{align}
\end{subequations}
\item the one-parameter class of solutions
\begin{subequations}
\begin{align}
&h_1=0,\\
&h_2=\frac{1}{e^{c_1-\frac{2}{\epsilon }} (1-\epsilon
   )^{w_m+1} \epsilon ^{-w_m-1}+1},
\end{align}	
\end{subequations} 
\item the 2-parameter class of solutions 
\begin{subequations}
\begin{align}
&h_1=
   1\pm\sqrt{\frac{e^{2 c_2} \epsilon ^{2/3}
   (1-\epsilon
   )^{w_m+\frac{1}{3}}}{e^{c_1}
   (1-\epsilon )^{w_m+1}+e^{2/\epsilon }
   \epsilon ^{w_m+1}}+1}, \\
&h_2=
   \frac{1}{e^{c_1-\frac{2}{\epsilon }}
   (1-\epsilon )^{w_m+1} \epsilon
   ^{-w_m-1}+1}.	
\end{align}
\end{subequations}  
\end{enumerate}

From all the above solutions the only one that satisfies the conditions $h_1(0)=0, 
h_2(0)=0, h_1'(0) =0, h_2'(0)=0$ is the trivial one. Thus, the dynamics on the center 
manifold is given by 
\begin{equation}
\frac{d\epsilon}{d\check{\tau}}= 3(1-\epsilon)\epsilon^2,
\end{equation}
that corresponds to a gradient-like equation for the potential $U(\epsilon)= 1/4 
\epsilon^3 (-4 + 3 \epsilon)$ for which the origin is a local maximum. Thus, the center 
manifold of the origin is unstable, and hence $R_{22}$ is a saddle. Integrating out the 
above equation we extract the solution 
\begin{equation}
\check{\tau} (\epsilon )= c_1-\frac{1}{3 \epsilon
   }-\frac{2}{3} \tanh ^{-1}(1-2 \epsilon ), 
\end{equation}
for which the origin is approached as $\check{\tau}\rightarrow -\infty$ and not as 
$\check{\tau}\rightarrow +\infty$, that indeed confirms the above statements.
\end{appendix}


\begin{thebibliography}{99}

 
 
 
\bibitem{Copeland:2006wr}
  E.~J.~Copeland, M.~Sami and S.~Tsujikawa,
     {\it{Dynamics of dark energy}},
  Int.\ J.\ Mod.\ Phys.\  D {\bf 15}, 1753 (2006),
[\href{http://xxx.lanl.gov/abs/hep-th/0603057}
{{\tt arXiv:hep-th/0603057}}].

 
 

\bibitem{Cai:2009zp}
  Y.~F.~Cai, E.~N.~Saridakis, M.~R.~Setare and J.~Q.~Xia,
     {\it{Quintom Cosmology: Theoretical implications and observations}},
  Phys.\ Rept.\  {\bf 493}, 1 (2010),
[\href{http://xxx.lanl.gov/abs/0909.2776}
{{\tt arXiv:0909.2776}}].

 
 
\bibitem{Olive:1989nu} 
  K.~A.~Olive,
     {\it{Inflation}},
  Phys.\ Rept.\  {\bf 190}, 307 (1990).

 
\bibitem{Bartolo:2004if} 
  N.~Bartolo, E.~Komatsu, S.~Matarrese and A.~Riotto,
     {\it{Non-Gaussianity from inflation: Theory and observations}},
  Phys.\ Rept.\  {\bf 402}, 103 (2004)
  [\href{http://xxx.lanl.gov/abs/astro-ph/0406398}
{{\tt arXiv:astro-ph/0406398}}].

 
 
 
\bibitem{Nojiri:2006ri}
  S.~Nojiri and S.~D.~Odintsov,
      {\it{Introduction to modified gravity and gravitational alternative
for dark
  energy}},
  eConf {\bf C0602061}, 06 (2006), Int.\ J.\ Geom.\ Meth.\ Mod.\ Phys.\ 
{\bf 4}, 115 (2007),
[\href{http://xxx.lanl.gov/abs/hep-th/0601213}
{{\tt arXiv:hep-th/0601213}}].


\bibitem{Capozziello:2011et}
  S.~Capozziello and M.~De Laurentis,
  {\it{Extended Theories of Gravity}},
  Phys.\ Rept.\  {\bf 509}, 167 (2011)
[\href{http://xxx.lanl.gov/abs/1108.6266}
{{\tt arXiv:1108.6266}}].
 
 
 
 
  \bibitem {Stelle:1976gc}
  K.~S.~Stelle,
  \textit{{Renormalization of Higher
Derivative Quantum Gravity}},
Phys.\ Rev.\ D \textbf{16}, 953 (1977).
 
\bibitem{Biswas:2011ar} 
  T.~Biswas, E.~Gerwick, T.~Koivisto and A.~Mazumdar,
  {\it{Towards singularity and ghost free theories of gravity}},
  Phys.\ Rev.\ Lett.\  {\bf 108}, 031101 (2012)
           [\href{http://xxx.lanl.gov/abs/1110.5249}
 {{\tt arXiv:1110.5249}}].

  

\bibitem{DeFelice:2010aj}
  A.~De Felice and S.~Tsujikawa,
  {\it{f(R) theories}},
  Living Rev.\ Rel.\  {\bf 13}, 3 (2010)
         [\href{http://xxx.lanl.gov/abs/1002.4928}
 {{\tt arXiv:1002.4928}}].

  
 

  
    
\bibitem{Nojiri:2010wj} 
  S.~'i.~Nojiri and S.~D.~Odintsov,
   {\it{Unified cosmic history in modified gravity: from F(R) theory to Lorentz
 non-invariant models}},
  Phys.\ Rept.\  {\bf 505}, 59 (2011)
          [\href{http://xxx.lanl.gov/abs/1011.0544}
 {{\tt arXiv:1011.0544}}].
 
 
 
  
  
\bibitem{Capozziello:2005ku} 
  S.~Capozziello, V.~F.~Cardone and A.~Troisi,
  {\it{Reconciling dark energy models with f(R) theories}},
  Phys.\ Rev.\ D {\bf 71}, 043503 (2005)
         [\href{http://xxx.lanl.gov/abs/astro-ph/0501426}
 {{\tt arXiv:astro-ph/0501426}}].
 
 
  
  
\bibitem{Amarzguioui:2005zq} 
  M.~Amarzguioui, O.~Elgaroy, D.~F.~Mota and T.~Multamaki,
   {\it{Cosmological constraints on f(r) gravity theories within the palatini 
approach}},
  Astron.\ Astrophys.\  {\bf 454}, 707 (2006)
           [\href{http://xxx.lanl.gov/abs/astro-ph/0510519}
 {{\tt arXiv:astro-ph/0510519}}].
 
 
  
\bibitem{Nojiri:2006gh} 
  S.~Nojiri and S.~D.~Odintsov,
  {\it{Modified f(R) gravity consistent with realistic cosmology: From matter 
dominated epoch to dark energy universe}},
  Phys.\ Rev.\ D {\bf 74}, 086005 (2006)
             [\href{http://xxx.lanl.gov/abs/hep-th/0608008}
 {{\tt arXiv:hep-th/0608008}}].
 
 
 
 


\bibitem{Nojiri:2005jg}
  S.~'i.~Nojiri and S.~D.~Odintsov,
   {\it{Modified Gauss-Bonnet theory as gravitational alternative for dark
 energy}},
  Phys.\ Lett.\ B {\bf 631}, 1 (2005)
                [\href{http://xxx.lanl.gov/abs/hep-th/0508049}
 {{\tt arXiv:hep-th/0508049}}].
 
 
 


\bibitem{DeFelice:2008wz}
  A.~De Felice and S.~Tsujikawa,
   {\it{Construction of cosmologically viable f(G) dark energy models}},
  Phys.\ Lett.\ B {\bf 675}, 1 (2009)
     [\href{http://xxx.lanl.gov/abs/0810.5712}
{{\tt arXiv:0810.5712}}].
 
 
\bibitem{Lovelock:1971yv}
  D.~Lovelock,
   {\it{The Einstein tensor and its generalizations}},
  J.\ Math.\ Phys.\  {\bf 12}, 498 (1971).




\bibitem{Deruelle:1989fj}
  N.~Deruelle and L.~Farina-Busto,
   {\it{The Lovelock Gravitational Field Equations in Cosmology}},
  Phys.\ Rev.\ D {\bf 41}, 3696 (1990).



\bibitem{Mannheim:1988dj}
  P.~D.~Mannheim and D.~Kazanas,
   {\it{Exact Vacuum Solution to Conformal Weyl Gravity and Galactic Rotation
Curves}},
  Astrophys.\ J.\  {\bf 342}, 635 (1989).

  
 \bibitem{Flanagan:2006ra}
  E.~E.~Flanagan,
    {\it{Fourth order Weyl gravity}},
  Phys.\ Rev.\ D {\bf 74}, 023002 (2006)
       [\href{http://xxx.lanl.gov/abs/astro-ph/0605504}
{{\tt arXiv:astro-ph/0605504}}].



 
  
\bibitem{Horava:2008ih}
P.~Horava, \textit{{Membranes at Quantum Criticality}}, JHEP
\textbf{0903}, 020 (2009)
[\href{http://xxx.lanl.gov/abs/0812.4287}{\texttt{arXiv:0812.4287}}].



 \bibitem{Kiritsis:2009sh}
  E.~Kiritsis and G.~Kofinas, \textit{{Horava-Lifshitz
Cosmology}}, Nucl.\ Phys.\ B \textbf{821}, 467 (2009)
[\href{http://xxx.lanl.gov/abs/0904.1334}{\texttt{arXiv:0904.1334}}].



 \bibitem{Saridakis:2012ui} 
 E.~N.~Saridakis, \textit{{Constraining
Horava-Lifshitz gravity from neutrino speed experiments}},
Gen.\ Rel.\ Grav.\ \textbf{45}, 387 (2013)
[\href{http://xxx.lanl.gov/abs/1110.0697}{\texttt{arXiv:1110.0697}}].
  
 
\bibitem{Nicolis:2008in} 
  A.~Nicolis, R.~Rattazzi and E.~Trincherini,
   {\it{The Galileon as a local modification of gravity}},
  Phys.\ Rev.\ D {\bf 79}, 064036 (2009)
         [\href{http://xxx.lanl.gov/abs/0811.2197}
{{\tt arXiv:0811.2197}}].

\bibitem{Deffayet:2009wt} 
  C.~Deffayet, G.~Esposito-Farese and A.~Vikman,
   {\it{Covariant Galileon}},
  Phys.\ Rev.\ D {\bf 79}, 084003 (2009)
           [\href{http://xxx.lanl.gov/abs/0901.1314}
{{\tt arXiv:0901.1314}}].

  
 
 \bibitem{Deffayet:2009mn} 
  C.~Deffayet, S.~Deser and G.~Esposito-Farese,
   {\it{Generalized Galileons: All scalar models whose curved background 
extensions
 maintain second-order field equations and stress-tensors}},
  Phys.\ Rev.\ D {\bf 80}, 064015 (2009)
           [\href{http://xxx.lanl.gov/abs/0906.1967}
{{\tt arXiv:0906.1967}}].


\bibitem{Leon:2012mt} 
  G.~Leon and E.~N.~Saridakis,
  {\it{Dynamical analysis of generalized Galileon cosmology}},
  JCAP {\bf 1303}, 025 (2013)
             [\href{http://xxx.lanl.gov/abs/1211.3088}
{{\tt arXiv:1211.3088}}].


 
  
\bibitem{deRham:2010kj}
  C.~de Rham, G.~Gabadadze and A.~J.~Tolley,
 {\it{Resummation of Massive Gravity}},
  Phys.\ Rev.\ Lett.\  {\bf 106}, 231101 (2011)
[\href{http://xxx.lanl.gov/abs/1011.1232}
{{\tt arXiv:1011.1232}}].

\bibitem{Hinterbichler:2011tt}
  K.~Hinterbichler,
 {\it{Theoretical Aspects of Massive Gravity}},
  Rev.\ Mod.\ Phys.\  {\bf 84}, 671 (2012)
[\href{http://xxx.lanl.gov/abs/1105.3735}
{{\tt arXiv:1105.3735}}].

 \bibitem{deRham:2014zqa} 
  C.~de Rham,
   {\it{Massive Gravity}},
  Living Rev.\ Rel.\  {\bf 17}, 7 (2014)
  [\href{http://xxx.lanl.gov/abs/1401.4173}
{{\tt arXiv:1401.4173}}].


\bibitem{Leon:2013qh} 
  G.~Leon, J.~Saavedra and E.~N.~Saridakis,
   {\it{Cosmological behavior in extended nonlinear massive gravity}},
  Class.\ Quant.\ Grav.\  {\bf 30}, 135001 (2013)
    [\href{http://xxx.lanl.gov/abs/1301.7419}
{{\tt arXiv:1301.7419}}].


 
 
 \bibitem{Ben09}
  G.~R.~Bengochea and R.~Ferraro,
  {\it{Dark torsion as the cosmic speed-up}},
  Phys.\ Rev.\ D \textbf{79}, 124019 (2009)
              [\href{http://xxx.lanl.gov/abs/0812.1205}
{{\tt arXiv:0812.1205}}].

  
 


\bibitem{Linder:2010py}
  E.~V.~Linder,
  {\it{Einstein's Other Gravity and the Acceleration of the
Universe}},
  Phys.\ Rev.\ D \textbf{81}, 127301 (2010)
                [\href{http://xxx.lanl.gov/abs/1005.3039}
{{\tt arXiv:1005.3039}}].

 
  
\bibitem{Chen:2010va}
  S.~H.~Chen, J.~B.~Dent, S.~Dutta and E.~N.~Saridakis,
     {\it{Cosmological perturbations in f(T) gravity}},
  Phys.\ Rev.\  D {\bf 83}, 023508 (2011)
 [\href{http://xxx.lanl.gov/abs/1008.1250}
{{\tt arXiv:1008.1250}}].
   
\bibitem{Cai:2011tc} 
  Y.~F.~Cai, S.~H.~Chen, J.~B.~Dent, S.~Dutta and E.~N.~Saridakis,
    {\it{Matter Bounce Cosmology with the f(T) Gravity}},
  Class.\ Quant.\ Grav.\  {\bf 28}, 215011 (2011)
   [\href{http://xxx.lanl.gov/abs/1104.4349}
{{\tt arXiv:1104.4349}}].
  
 
  
\bibitem{Kofinas:2014owa}
  G.~Kofinas and E.~N.~Saridakis,
  {\it{Teleparallel equivalent of Gauss-Bonnet gravity and its
modifications}},
  Phys.\ Rev.\ D {\bf 90}, no. 8, 084044 (2014)
     [\href{http://xxx.lanl.gov/abs/1404.2249}
{{\tt arXiv:1404.2249}}].
 
  
\bibitem{Kofinas:2014aka} 
  G.~Kofinas, G.~Leon and E.~N.~Saridakis,
 {\it{Dynamical behavior in $f(T,T_G)$ cosmology}},
  Class.\ Quant.\ Grav.\  {\bf 31}, 175011 (2014),
       [\href{http://xxx.lanl.gov/abs/1404.7100}
{{\tt arXiv:1404.7100}}].
 
 
\bibitem{Kofinas:2014daa} 
  G.~Kofinas and E.~N.~Saridakis,
  {\it{Cosmological applications of $F(T,T_G)$ gravity}},
  Phys.\ Rev.\ D {\bf 90}, no. 8, 084045 (2014)
         [\href{http://xxx.lanl.gov/abs/1408.0107}
{{\tt arXiv:1408.0107}}].

 
 
   
\bibitem{Cortes:2015ola} 
  M.~Cortes, H.~Gomes and L.~Smolin,
   {\it{Time asymmetric extensions of general relativity}},
      [\href{http://xxx.lanl.gov/abs/1503.06085}
 {{\tt arXiv:1503.06085}}].
  
  
  \bibitem{Coley:2003mj}
A.~A. Coley.
  {\it{Dynamical systems and cosmology}},
\newblock Dordrecht, Netherlands: Kluwer (2003).


  \bibitem{Leon2011} G. Leon and C. R. Fadragas, {\it{Cosmological Dynamical
Systems}}, LAP LAMBERT Academic Publishing,
(2011),
 [\href{http://xxx.lanl.gov/abs/1412.5701}
 {{\tt arXiv:1412.5701}}].
 
 
\bibitem{Arnowitt:1962hi} 
  R.~L.~Arnowitt, S.~Deser and C.~W.~Misner,
  {\it{The Dynamics of general relativity}},
  Gen.\ Rel.\ Grav.\  {\bf 40}, 1997 (2008)
   [\href{http://xxx.lanl.gov/abs/gr-qc/0405109}
 {{\tt arXiv:gr-qc/0405109}}].
 
 

  
  
\bibitem{Gomes:2010fh} 
  H.~Gomes, S.~Gryb and T.~Koslowski,
  {\it{Einstein gravity as a 3D conformally invariant theory}},
  Class.\ Quant.\ Grav.\  {\bf 28}, 045005 (2011)
        [\href{http://xxx.lanl.gov/abs/1010.2481}
 {{\tt arXiv:1010.2481}}].
 


\bibitem{Gomes:2011zi} 
  H.~Gomes and T.~Koslowski,
  {\it{The Link between General Relativity and Shape Dynamics}},
  Class.\ Quant.\ Grav.\  {\bf 29}, 075009 (2012)
          [\href{http://xxx.lanl.gov/abs/1101.5974}
 {{\tt arXiv:1101.5974}}].
 
 
 
\bibitem{Gomes:2013naa} 
  H.~Gomes,
  {\it{Conformal geometrodynamics regained: gravity from duality}},
  Annals Phys.\  {\bf 355}, 224 (2015)
          [\href{http://xxx.lanl.gov/abs/1310.1699}
 {{\tt arXiv:1310.1699}}].
 
 
  

  
\bibitem{Ashtekar:1986yd} 
  A.~Ashtekar,
  {\it{New Variables for Classical and Quantum Gravity}},
  Phys.\ Rev.\ Lett.\  {\bf 57}, 2244 (1986).
  
\bibitem{Thurston:1982zz} 
  W.~P.~Thurston,
   {\it{Three dimensional manifolds, Kleinian groups and hyperbolic geometry}},
  Bull.\ Am.\ Math.\ Soc.\  {\bf 6}, 357 (1982).

  
  
  
\bibitem{Perko} 
L. Perko, 
{\it{Differential Equations and Dynamical Systems}}, Springer, Heidelberg (2006).


\bibitem{Ellis} 
{\it{Dynamical Systems in Cosmology}}, 
edited by J. Wainwright 
and
G. F. R. Ellis, Cambridge University Press, Cambridge (1997).


\bibitem{Copeland:1997et}
  E.~J.~Copeland, A.~R.~Liddle and D.~Wands,
     {\it{Exponential potentials and cosmological scaling solutions}},
  Phys.\ Rev.\  D {\bf 57}, 4686 (1998)
[\href{http://xxx.lanl.gov/abs/gr-qc/9711068}
{{\tt arXiv:gr-qc/9711068}}].


\bibitem{Ferreira:1997au}
  P.~G.~Ferreira and M.~Joyce,
     {\it{Structure formation with a self-tuning scalar field}},
  Phys.\ Rev.\ Lett.\  {\bf 79}, 4740 (1997)
[\href{http://xxx.lanl.gov/abs/astro-ph/9707286}
{{\tt arXiv:astro-ph/9707286}}].


\bibitem{Chen:2008ft}
  X.~m.~Chen, Y.~g.~Gong and E.~N.~Saridakis,
     {\it{Phase-space analysis of interacting phantom cosmology}},
  JCAP {\bf 0904}, 001 (2009)
[\href{http://xxx.lanl.gov/abs/0812.1117}
{{\tt arXiv:0812.1117}}].


\bibitem{Cotsakis:2013zha}
  S.~Cotsakis and G.~Kittou,
  {\it{Flat limits of curved interacting cosmic fluids}},
  Phys.\ Rev.\ D {\bf 88}, 083514 (2013)
[\href{http://xxx.lanl.gov/abs/1307.0377}
{{\tt arXiv:1307.0377}}].


\bibitem{Giambo':2009cc}
  R.~Giambo and J.~Miritzis,
  {\it{Energy exchange for homogeneous and isotropic universes with a scalar 
field coupled to matter}},
  Class.\ Quant.\ Grav.\  {\bf 27} (2010) 095003
  [\href{http://xxx.lanl.gov/abs/0908.3452}
{{\tt arXiv:0908.3452}}].

 \bibitem{Xu:2012jf}
   C.~Xu, E.~N.~Saridakis and G.~Leon,
  {\it{Phase-Space analysis of Teleparallel Dark Energy}},
  JCAP {\bf 1207}, 005 (2012)
     [\href{http://xxx.lanl.gov/abs/1202.3781}
 {{\tt arXiv:1202.3781}}].

\bibitem{Leon:2009ce} 
  G.~Leon, Y.~Leyva, E.~N.~Saridakis, O.~Martin and R.~Cardenas,
  {\it{Falsifying Field-based Dark Energy Models}},  In: {\it{Dark Energy: Theory, 
Developements, 
and Implications}}. Nova Science Publishing, New York (2010), 
[\href{http://xxx.lanl.gov/abs/0912.0542}
{{\tt arXiv:0912.0542}}].


 
   
\bibitem{Coley:2000yc} 
  A.~Coley and M.~Goliath,
   {\it{Closed cosmologies with a perfect fluid and a scalar field}},
  Phys.\ Rev.\ D {\bf 62}, 043526 (2000)
  [\href{http://xxx.lanl.gov/abs/gr-qc/0004060}
{{\tt arXiv:gr-qc/0004060}}].

 

\bibitem{Goliath:1998na} 
  M.~Goliath and G.~F.~R.~Ellis,
   {\it{Homogeneous cosmologies with cosmological constant}},
  Phys.\ Rev.\ D {\bf 60}, 023502 (1999)
    [\href{http://xxx.lanl.gov/abs/gr-qc/9811068}
{{\tt arXiv:gr-qc/9811068}}].

 
  
  
\bibitem{Halliwell:1986ja} 
  J.~J.~Halliwell,
   {\it{Scalar Fields in Cosmology with an Exponential Potential}},
  Phys.\ Lett.\ B {\bf 185}, 341 (1987).

\bibitem{vandenHoogen:1999qq} 
  R.~J.~van den Hoogen, A.~A.~Coley and D.~Wands,
   {\it{Scaling solutions in Robertson-Walker space-times}},
  Class.\ Quant.\ Grav.\  {\bf 16}, 1843 (1999)
      [\href{http://xxx.lanl.gov/abs/gr-qc/9901014}
{{\tt arXiv:gr-qc/9901014}}].

 
 
 
  
  
  
  
\bibitem{Alho:2015cza} 
A.~Alho, J.~Hell and C.~Uggla,
{\it{Global dynamics and asymptotics for monomial scalar field potentials and perfect 
fluids}},
[\href{http://xxx.lanl.gov/abs/1503.06994}
{{\tt arXiv:1503.06994}}].

 
  

\bibitem{wiggins} S. Wiggins,
 {\it{ Introduction to Applied Nonlinear Dynamical Systems and Chaos}},
Springer, New York (2003). 

\bibitem{Copeland:2009be} 
  E.~J.~Copeland, S.~Mizuno and M.~Shaeri,
 {\it{Dynamics of a scalar field in Robertson-Walker spacetimes}},
  Phys.\ Rev.\ D {\bf 79}, 103515 (2009)
  [\href{http://xxx.lanl.gov/abs/0904.0877}
{{\tt arXiv:0904.0877}}].

 


\bibitem{Planck:2015xua} 
  P.~A.~R.~Ade {\it et al.}  [Planck Collaboration],
{\it{Planck 2015 results. XIII. Cosmological parameters}},
[\href{http://xxx.lanl.gov/abs/1502.01589}
{{\tt arXiv:1502.01589}}].

\bibitem{Caldwell:2003vq} 
  R.~R.~Caldwell, M.~Kamionkowski and N.~N.~Weinberg,
{\it{Phantom energy and cosmic doomsday}},
  Phys.\ Rev.\ Lett.\  {\bf 91}, 071301 (2003)
  [\href{http://xxx.lanl.gov/abs/astro-ph/0302506}
{{\tt arXiv:astro-ph/0302506}}].
 

\bibitem{Sami:2003xv} 
  M.~Sami and A.~Toporensky,
  {\it{Phantom field and the fate of universe}},
  Mod.\ Phys.\ Lett.\ A {\bf 19}, 1509 (2004)
    [\href{http://xxx.lanl.gov/abs/gr-qc/0312009}
{{\tt arXiv:gr-qc/0312009}}].
 
  
\bibitem{Hao:2004ky}
J.~G.~Hao and X.~Z.~Li,
{\it{Generalized quartessence cosmic dynamics: Phantom or quintessence
with de Sitter attractor}},
Phys.\ Lett.\ B {\bf 606}, 7 (2005)
          [\href{http://xxx.lanl.gov/abs/astro-ph/0404154}
{{\tt arXiv:astro-ph/0404154}}].

 

\bibitem{Nojiri:2015fia} 
  S.~Nojiri, S.~D.~Odintsov, V.~K.~Oikonomou and E.~N.~Saridakis,
{\it{Singular cosmological evolution using canonical and phantom scalar fields}},
          [\href{http://xxx.lanl.gov/abs/1503.08443}
{{\tt arXiv:1503.08443}}].


 
\bibitem{Nojiri:2005sx}
S.~Nojiri, S.~D.~Odintsov and S.~Tsujikawa,
{\it{Properties of singularities in (phantom) dark energy
universe}},
Phys.\ Rev.\ D {\bf 71}, 063004 (2005)
   [\href{http://xxx.lanl.gov/abs/hep-th/0501025}
{{\tt arXiv:hep-th/0501025}}].

  
  
\bibitem{Longo:1987ub}
  M.~J.~Longo,
{\it{Tests of relativity from SN1987a}},
  Phys.\ Rev.\  D {\bf 36}, 3276 (1987).

\bibitem{Hirata:1987hu}
  K.~Hirata {\it et al.}  [KAMIOKANDE-II Collaboration],
{\it{Observation of a Neutrino Burst from the Supernova SN 1987a}},
  Phys.\ Rev.\ Lett.\  {\bf 58}, 1490 (1987).

\bibitem{Bionta:1987qt}
  R.~M.~Bionta {\it et al.},
 {\it{Observation of a Neutrino Burst in Coincidence with Supernova SN 1987a
 in
  the Large Magellanic Cloud}},
  Phys.\ Rev.\ Lett.\  {\bf 58}, 1494 (1987).
  


\end{thebibliography}
\end{document}